\documentclass[10pt,conference]{IEEEtran}

\IEEEoverridecommandlockouts

\usepackage{cite}
\usepackage{tabularx}

\newcommand\clearrow{\global\let\rowmac\relax}
\clearrow
\usepackage{adjustbox}
\usepackage[flushleft]{threeparttable}
\usepackage{amsmath, amsfonts}
\usepackage{algorithmic}
\usepackage{graphicx}
\usepackage{subfig}
\usepackage{textcomp}
\usepackage{xcolor}
\usepackage{xspace}
\usepackage{booktabs}
\usepackage{enumitem}
\usepackage{microtype}
\usepackage{multirow}
\usepackage{colortbl}
\usepackage{rotating}
\usepackage{pifont}
\usepackage{wrapfig}

\usepackage{svg}

\usepackage{balance}

\usepackage{hyperref}
\hypersetup{
    colorlinks=true,    
    linkcolor=black,    
    citecolor=black,     
    filecolor=magenta,  
    urlcolor=black       
}

\definecolor{myorange}{HTML}{C55A11}  

\def\BibTeX{{\rm B\kern-.05em{\sc i\kern-.025em b}\kern-.08em
    T\kern-.1667em\lower.7ex\hbox{E}\kern-.125emX}}

\begin{document}

\title{Seeing is Fixing: Cross-Modal Reasoning with Multimodal LLMs for Visual Software Issue Fixing}

\author{
\IEEEauthorblockN{
Kai Huang\IEEEauthorrefmark{1},
Jian Zhang\IEEEauthorrefmark{2}\thanks{\IEEEauthorrefmark{2}Corresponding author: Jian Zhang (jian\_zhang@ntu.edu.sg).},
Xiaofei Xie\IEEEauthorrefmark{3},
Chunyang Chen\IEEEauthorrefmark{1}
}

\IEEEauthorblockA{
\IEEEauthorrefmark{1}Technical University of Munich,
\IEEEauthorrefmark{2}Nanyang Technological University,
\IEEEauthorrefmark{3}Singapore Management University
}

\IEEEauthorblockA{
\{kai-kevin.huang, chun-yang.chen\}@tum.de,
jian\_zhang@ntu.edu.sg, xfxie@smu.edu.sg
}
}

\maketitle

\begin{abstract}

Large language model (LLM)-based automated program repair (APR) techniques have shown promising results in resolving real-world github issue tasks. Existing APR systems are primarily evaluated in unimodal settings (e.g., SWE-bench), relying solely on textual issue descriptions and source code.
However, these autonomous systems struggle to resolve multimodal problem scenarios (e.g., SWE-bench M) due to limitations in interpreting and leveraging visual information.
In multimodal scenarios, LLMs need to rely on visual information in the graphical user interface (GUI) to understand bugs and generate fixes.
To bridge this gap,
we propose GUIRepair, a cross-modal reasoning approach for resolving multimodal issue scenarios by understanding and capturing visual information. 
Specifically, GUIRepair integrates two key components, Image2Code and Code2Image—to enhance fault comprehension and patch validation.
Image2Code extracts relevant project documents based on the issue report, then applies this domain knowledge to generate the reproduced code responsible for the visual symptoms, effectively translating GUI images into executable context for better fault comprehension.
Code2Image replays the visual issue scenario using the reproduced code and captures GUI renderings of the patched program to assess whether the fix visually resolves the issue, providing feedback for patch validation.
We evaluate GUIRepair on SWE-bench M, and the approach demonstrates significant effectiveness. 
When utilizing GPT-4o as the base model, GUIRepair solves 157 instances, outperforming the best open-source baseline by 26 instances. 
Furthermore, when using o4-mini as the base model, GUIRepair can achieve even better results and solve 175 instances, outperforming the top commercial system by 22 instances.
This emphasizes the success of our new perspective on incorporating cross-modal reasoning by understanding and capturing visual information to 
resolve multimodal issues.

\end{abstract}


\maketitle

\section{Introduction}

\begin{figure}[t]
    \centering
    {\includegraphics[width=1.0\linewidth]{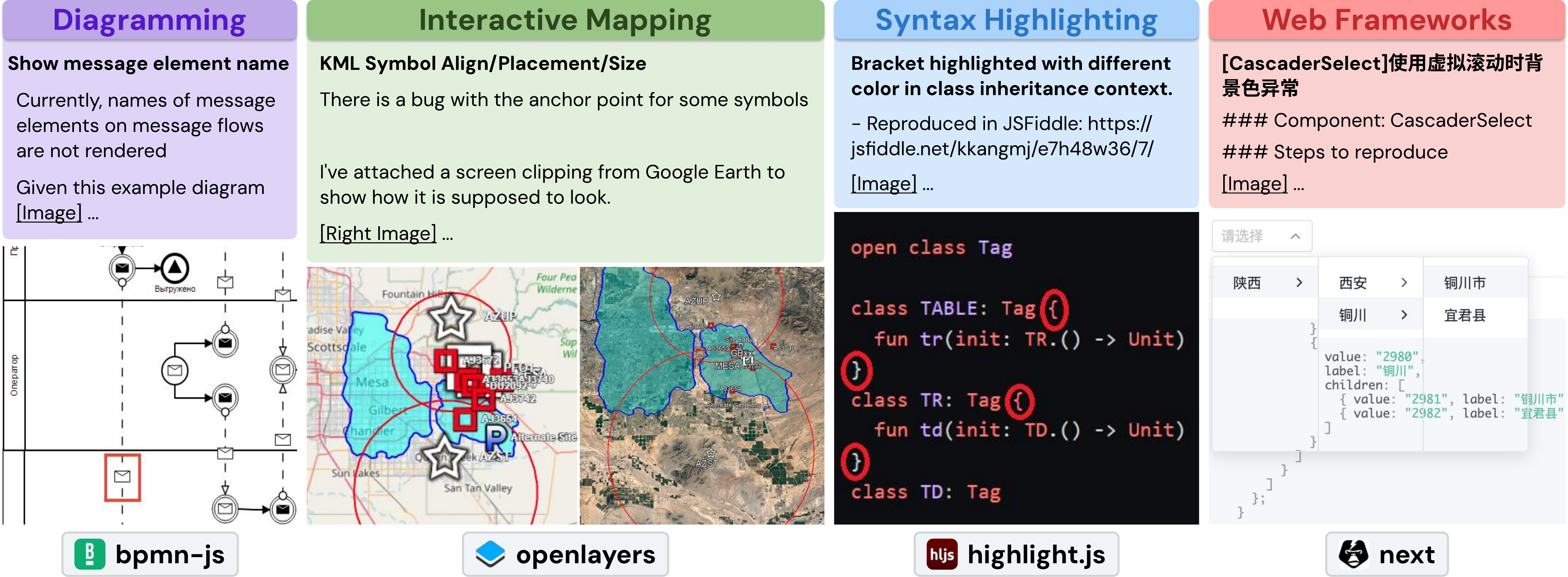}}
    \caption{Four visual task instances from SWE-bench M~\cite{SWEBenchM}.}
    \label{fig:instances_example}
    \vspace{-15px}
\end{figure}

Automated program repair (APR) techniques \cite{LLM4APR_Report,APR_CACM,APR_Survey_CSUR,APR_Survey_TOSEM,APR_Survey_Arxiv,APR_Survey_Renzullo} aim to automatically resolve software defects, reducing manual effort and improving software quality~\cite{APR_industry}. With the advent of the large language model (LLM) era, LLM-driven APR techniques~\cite{APR_Survey_Arxiv} show promising results in the software engineering (SE) community~\cite{LLM4SE_survey,LLM4SE_Survey_Arxiv}. 
Researchers have proposed many autonomous systems to repair software bugs~\cite{RepairAgent,FixAgent,ChatRepair,ThinkRepair,Contrastrepair}.
Recently, the APR community has begun to focus on a new challenge, SWE-bench~\cite{SWEBench}, designed to evaluate LLMs in solving real-world GitHub issue instances~\cite{AutoCodeRover,SWE-agent,Agentless,SpecRover,LingmaSWE,LingmaAgent}.
These issues are typically represented by textual components including natural language and programming language, few of which contain an image (i.e., screenshot).
However, many real-world software engineering tasks involve visual assets, and widely-used front-end libraries enable users to build applications with graphical user interfaces (GUIs). As shown in Figure~\ref{fig:instances_example}, this is especially common in domains such as web development and interactive application design. When bugs occur in such applications, users often report issues in the repository, accompanied by both textual descriptions and screenshots. Developers maintaining these libraries rely not only on the code and issue descriptions but also heavily on visual cues from the screenshots to understand and diagnose the problems. Without visual context, it becomes difficult to identify the root cause of the issue or to verify whether it has been  resolved.

Unfortunately, there has been limited research on automated repair for visual software issues, despite their practical significance. Existing approaches~\cite{Agentless,SWE-agent,ChatRepair,RepairAgent,FixAgent,Adversarial,ThinkRepair,Contrastrepair,AutoCodeRover,SpecRover,LingmaSWE,LingmaAgent} primarily target unimodal scenarios that involve only textual information, such as code and natural language. These methods are typically evaluated on benchmarks like SWE-bench~\cite{SWEBench}, Defects4J~\cite{Defects4J}, and HumanEval~\cite{LLM4APR_Jiang}, which are not designed to address multimodal tasks involving both visual and textual elements~\cite{SWEBenchM}. Recent findings~\cite{SWEBenchM} reveal that top-performing systems on unimodal benchmarks struggle significantly when applied to multimodal settings. For example, SWE-agent~\cite{SWE-agent}, which performs strongly on SWE-bench, successfully solves only 12\% of tasks in the multimodal SWE-bench M benchmark. This sharp performance decline highlights the unique challenges of multimodal scenarios and underscores the urgent need to advance existing APR systems to better handle visual problems.


In order to explore and reveal the reasons behind this, 
we analyze the workflows of existing AI systems and identify two key limitations that hinder their effectiveness in visual issues:

    \noindent \textbf{1) Insufficient Understanding of Visual Issues.}
    Most LLM-driven autonomous systems focus on analyzing textual issue descriptions and source code to identify and repair defects. However, they often overlook the relationship between visual outputs (e.g., screenshots) and the underlying code components.
    Directly incorporating images from issue reports poses challenges—LLMs typically lack the domain-specific context needed to interpret visual scenarios. They may struggle to identify which code components are responsible for the visual issue or how these components interact to produce the faulty behavior.
    Consequently, this limited understanding of visual symptoms can hinder accurate fault localization and reduce the effectiveness of generated patches.
    \noindent \textbf{2) Lack of Feedback for Repair Behaviors.}
    Existing autonomous systems typically rely on executing test suites to validate and select valid patches~\cite{RepairAgent,FixAgent,ChatRepair}. However, SWE-bench M omits test cases to prevent potential data leakage, removing a critical feedback signal for patch validation.
    While some approaches~\cite{Agentless,SWE-agent} generate test cases as substitutes, they are not well-suited for visual issues. 
    On the one hand, certain visual problems rely on pixel-level comparison in visualization-based test cases~\cite{SWEBenchM}, which are challenging for language models to generate accurately. On the other hand, although reproducing issue scenarios can provide feedback on patch effectiveness~\cite{SWE-agent,BugReproduce}, most multimodal task instances in SWE-bench M lack reproduction scripts~\cite{SWEBenchM}, making it difficult to recreate the necessary visual conditions. As a result, a key challenge is how to reliably reproduce visual scenarios to enable effective patch validation.
In this paper, we propose GUIRepair, a cross-modal transformation solution designed for multimodal scenarios. GUIRepair addresses the challenges of multimodal issue resolution by enabling seamless transformation between textual and visual information, thereby facilitating both the understanding and utilization of visual content.
Specifically, GUIRepair comprises two core components: Image2Code and Code2Image. For simplicity, the term \textit{image} is used to refer to a screenshot in this paper.
Image2Code enhances bug comprehension by interpreting visual information to support accurate fault localization, while Code2Image generates visual representations from code to provide feedback for patch validation.
\textbf{1) Image2Code}: Before performing fault localization, GUIRepair first selectively retrieves relevant project documentation based on the multimodal issue report to better understand the visual problem. It then uses these retrieved documents as domain knowledge within the 
knowledge injection
paradigm to generate the code that reproduces the visual issue. This process transforms high-level visual information into its corresponding code behavior, providing essential contextual knowledge to map the visual problem to specific code components. In essence, by reasoning from visual issue scenarios to their textual code representations, GUIRepair enables large language models to develop a more comprehensive understanding of the problem using both visual and textual features.
\textbf{2) Code2Image}: After generating patches, GUIRepair replays the bug using the previously generated reproduction code and captures the resulting behavior in the GUI. It then evaluates whether the visual presentation of the patched program aligns with the expected outcome described in the issue report, thereby filtering for potentially correct patches. This step provides critical feedback for patch validation by converting the abstract fixing behavior into concrete, observable visual effects.
That is, by translating textual repair behaviors into visual outcomes, GUIRepair enables language models to verify patch effectiveness more intuitively through multimodal cues.
Finally, these two core components are seamlessly integrated into the overall repair workflow including fault localization, patch generation and patch validation. This bidirectional transformation enables GUIRepair to reason across visual and programmatic modalities, significantly enhancing the accuracy and effectiveness of automated issue repair.

We evaluate GUIRepair on the SWE-bench M benchmark comparing against state-of-the-art autonomous systems. 
Our results show that GUIRepair is able to achieve the best performance (157 resolved instances) than other open-source approaches when keeping the same base model GPT-4o. 
In particular, GUIRepair’s two key components with Image2Code and Code2Image resolved 21 more instances than the basic repair workflow, marking a 15.44\% improvement. 
Furthermore, when using the advanced reasoning model o4-mini, GUIRepair can achieve amazing results (175 resolved instances), which solves 22 instances more than the best commercial system.

Overall, the main contributions of this paper are as follows:

\begin{itemize}[leftmargin=0.3cm]
    \item \textbf{Dimension.} To the best of our knowledge, this is the first work to explicitly address issues with visual cues in front-end libraries, which opens a new dimension in automated issue resolution by incorporating textual and visual modalities.

    \item \textbf{Technique.}
    We propose GUIRepair, a bidirectional cross-modal transformation approach to solve multimodal software development problems. GUIRepair performs image-to-code transformation to diagnose issues by extracting semantic features from GUI screenshots and mapping them to code-level representations. Furthermore, it applies code-to-image transformation to validate repair patches by rendering the updated GUI and assessing its visual correctness. 

    \item \textbf{Evaluation.}
    We conduct extensive experiments on the popular SWE-bench M benchmark, comparing GUIRepair against a range of open-source and commercial autonomous systems. GUIRepair resolves 157 issues, outperforming all existing approaches. Notably, its two key components (Image2Code and Code2Image) contribute a 15.44\% improvement in overall effectiveness, highlighting the impact of cross-modal reasoning in addressing complex multimodal program repair tasks.
    We have released our code and experimental data \cite{GUIRepair_link}.

\end{itemize}
\section{Motivation}
In this section, we 
analyze the gap of existing autonomous systems in resolving visual software problems, which motivates us to design a cross-modal reasoning approach. 




\subsection{Challenge 1: Fault Understanding.}
\label{sec:motivation_image2code}
\begin{figure}[t]
    \centering
    {\includegraphics[width=1.0\linewidth, trim=113 127.6 112 127.6, clip]{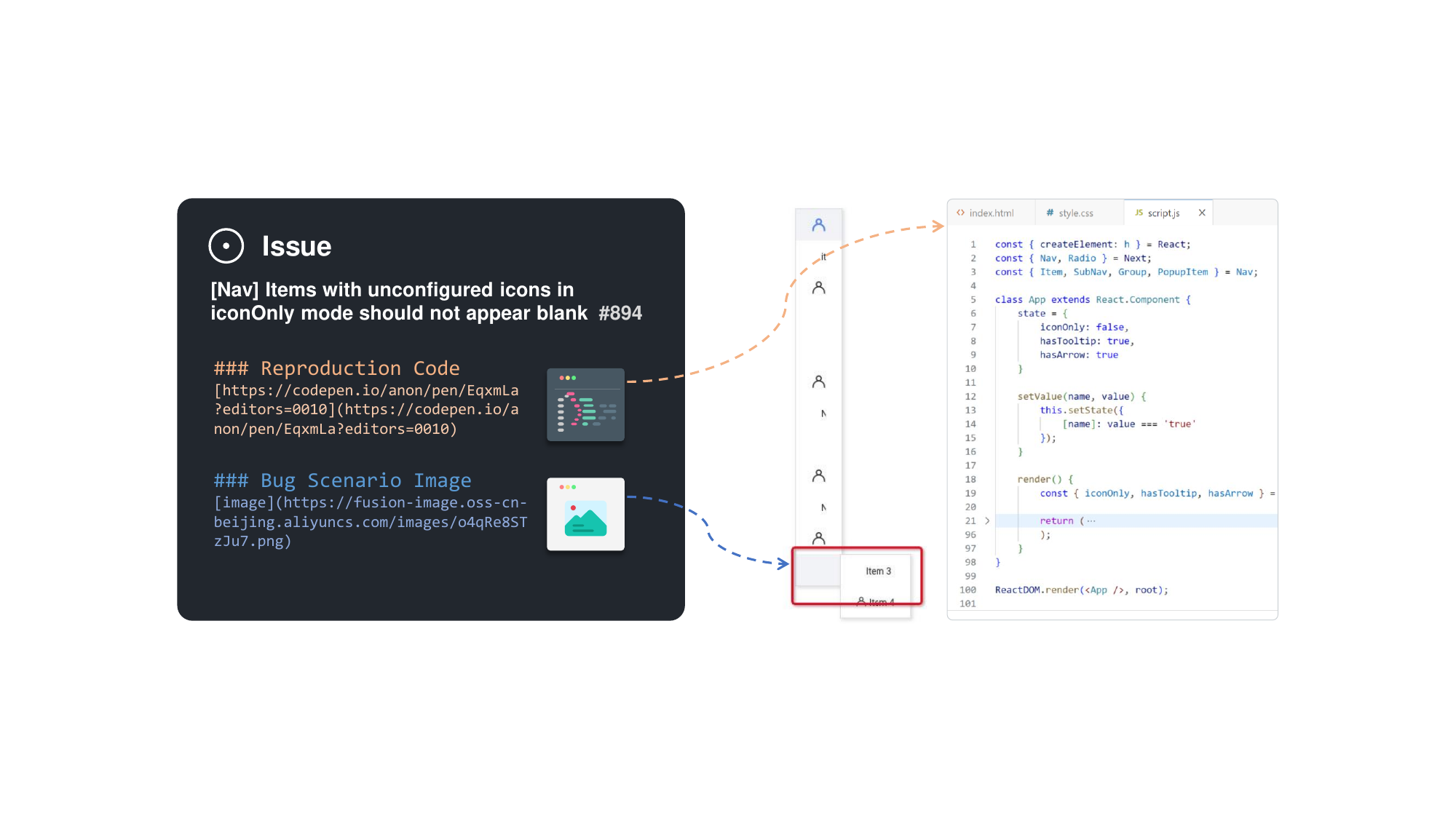}}
    \caption{A motivating example of alibaba-fusion/next-895~\cite{next-895}.} 
    \label{fig:example_image2code}
    \vspace{-8px}
\end{figure}

In unimodal scenarios, user-described issue symptoms are typically conveyed through text~\cite{SWEBench}. However, multimodal issue reports often involve both textual and visual elements~\cite{SWEBenchM}. As shown in Figure~\ref{fig:example_image2code}, these reports usually contain three layers of information: 1) natural language descriptions, 2) visual issue images, and 3) reproduced code that links symptoms to underlying components.
Text alone often fails to capture visual issues. A recent study~\cite{SWEBenchM} found that 80\% of visual symptoms cannot be fully described in text, and images are essential for issue-solving in 83.5\% of cases.

Even when text and images are available, models may still struggle to analyze visual symptoms of users' applications and identify root causes in the used front-end library. This is because LLMs may lack human expert knowledge about the specific front-end project~\cite{AdbGPT}, making it difficult for them to understand the code behavior behind the visual information. This gap stems from the lack of reproduced code, which clarifies the relationship between visual symptoms and code behavior. Yet, only 17\% of SWE-bench M reports include such code~\cite{SWEBenchM}, leaving most cases without sufficient context. This highlights a key challenge in multimodal bug resolution: inferring code behavior from visual symptoms for effective fault localization.

\subsection{Challenge 2: Patch Validation.}
\label{sec:motivation_code2image}
AI systems typically generate multiple candidate patches, but only the top-ranked one is selected in order to best meet developers’ needs (e.g., SWE-bench uses the Pass@1 metric). Therefore, accurately identifying which patches are effective is a crucial post-processing step. 
However, most current systems typically cannot perform patch validation on visual scenes directly~\cite{AgentlessLite}, limiting their ability to provide meaningful feedback on patch correctness. 

\begin{figure}[t]
    \centering
    {\includegraphics[width=1.0\linewidth, trim=79.5 114 79.5 114, clip]{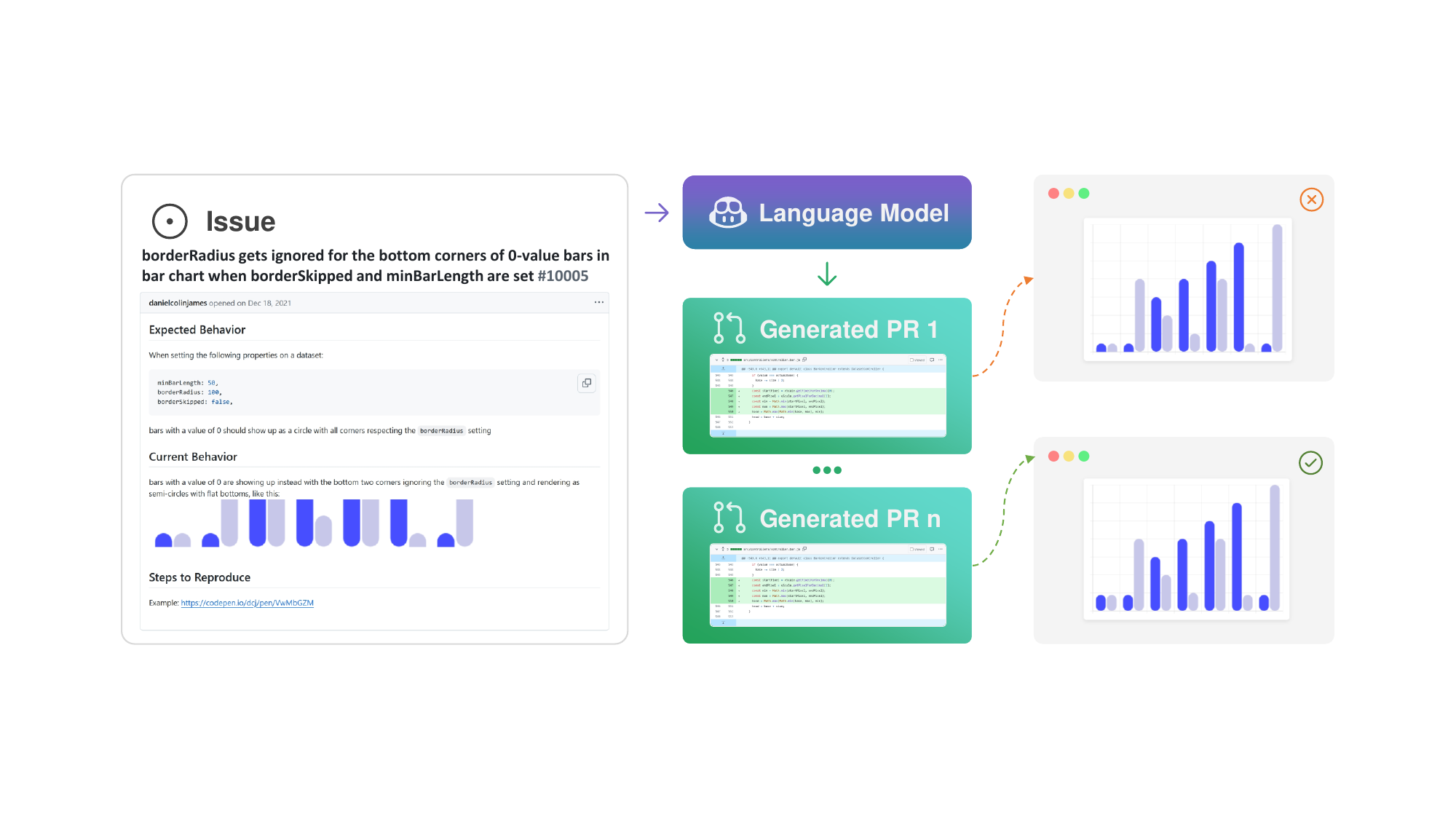}}
    \caption{A motivating example of chartjs/Chart.js-10157~\cite{chartjs-10157}.} 
    \label{fig:example_code2image}
    \vspace{-8px}
\end{figure}

While some approaches validate patches in unimodal settings using textual test cases~\cite{Agentless}, generating reliable visual oracles that involves pixel-level accuracy remains difficult, especially for issues requiring visual testing~\cite{Pixelmatch,Puppeteer,SWEBenchM}.
For example, data visualization and interactive mapping frameworks like \textit{Chart.js}\cite{chartjs} and \textit{OpenLayers}\cite{openLayers} often rely on pixel-level visual tests to verify correctness. As shown in Figure~\ref{fig:example_code2image}, we illustrate a \textit{Chart.js} issue~\cite{chartjs-10157}: \textit{borderRadius is ignored for the bottom corners of 0-value bars when \texttt{borderSkipped} and \texttt{minBarLength} are set}. Different proposed fixes (PR 1–N) produce subtle but distinct visual effects. Therefore, unlike in textual domains, generating such image test cases is infeasible for LLMs, as they cannot produce accurate visual references. This highlights a central challenge in visual patch validation: capturing and assessing visual effects as the repair feedback. 

\section{Approach}



\begin{figure*}[ht]
    \centering
    \includegraphics[width=1.0\linewidth, trim=24 10 20 10, clip]
    {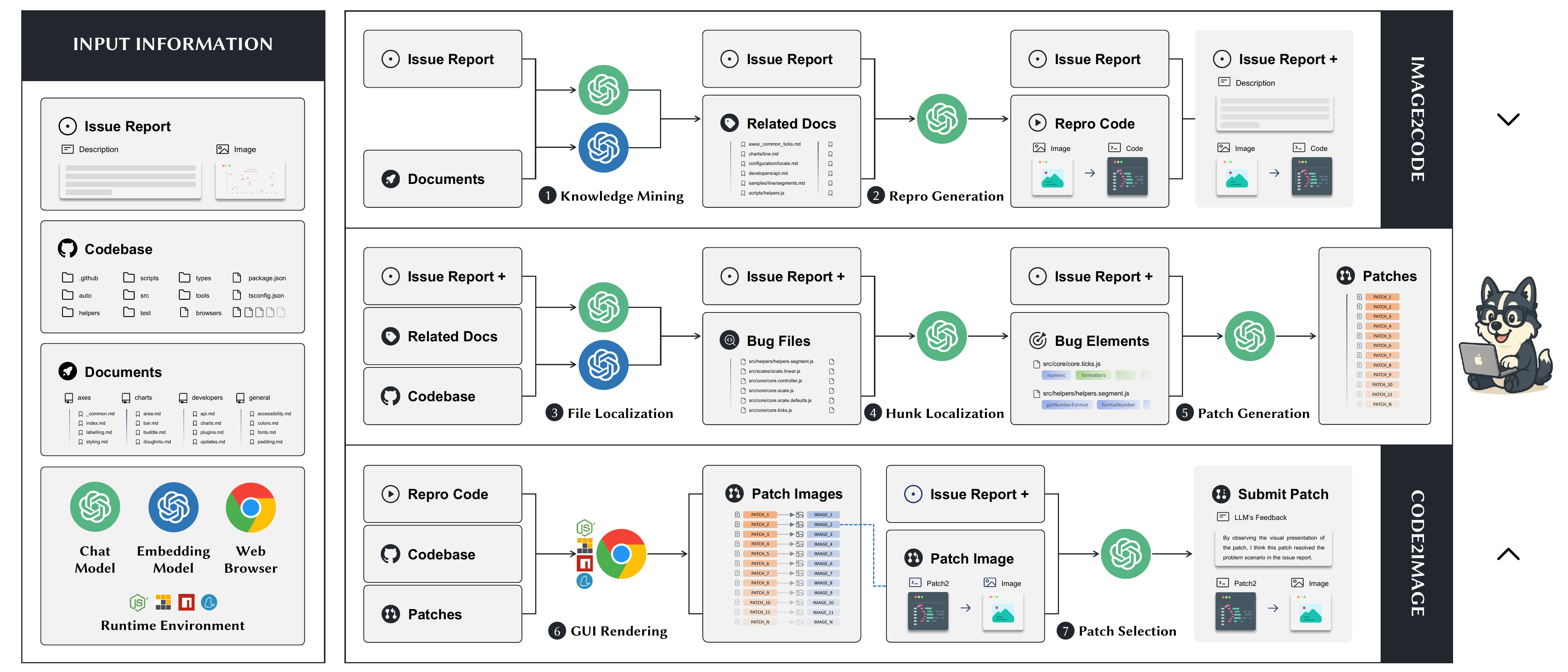}
    \caption{The overview of GUIRepair. 
    {\small Note: The left shows the system input, and the right illustrates the step-by-step repair workflow. 
    }
    } 
    \label{fig:guirepair_workflow}
    \vspace{-10px}
\end{figure*}

In this section, we introduce GUIRepair to solve visual software issues by reasoning the correlation between visual information from user-submitted reports and code components from the front-end library.
Basically, GUIRepair follows the traditional APR pipeline~\cite{APR_Survey_CSUR}, which is organized into four main steps: Fault Comprehension, Fault Localization, Patch Generation, and Patch Validation.
Figure~\ref{fig:guirepair_workflow} illustrates the overall workflow.
In the fault comprehension phase (\ding{182}–\ding{183}), GUIRepair first mines project-specific knowledge and generates reproduction code to help LLMs better understand the issue of the library. This reproducible code is essential for replicating the problem scenario in user applications.
In the fault localization phase (\ding{184}–\ding{185}), GUIRepair leverages the 
knowledge injection paradigm~\cite{knowledgeInjection_Survey}, 
utilizing the domain knowledge acquired in the previous phase to assist LLMs in accurately identifying the code components associated with specific visual elements.
During the patch generation phase (\ding{186}), GUIRepair prompts the LLM to perform multiple sampling rounds, producing candidate patches for the library based on the previously learned knowledge of the bug scenario.
Finally, in the patch validation phase (\ding{187}–\ding{188}), instead of generating test cases, GUIRepair captures the visual effects of generated patches by rendering GUI of the reproduced code and provides feedback to select the most promising patch.

\subsection{Fault Comprehension}
\label{sec:image2code}

As highlighted in Section~\ref{sec:motivation_image2code}, existing AI systems often struggle with multimodal software engineering (SWE) tasks due to limited understanding of the interplay between visual and code representations. This lack of multimodal comprehension impairs effective fault localization.
To address this challenge, GUIRepair introduces the Image2Code component, which enhances fault comprehension by enabling models to acquire project-specific domain knowledge and reason about the code components behind visual scenarios. As shown in Figure~\ref{fig:guirepair_workflow}, Image2Code comprises two key steps:
\ding{182} Knowledge Mining, and
\ding{183} Reproduction Generation.




\subsubsection{Knowledge Mining}
\label{sec:knowledge_mining}
\begin{wrapfigure}{r}{0.24\textwidth} 
    \vspace{-15px}
        \begin{center}
            \subfloat[Example of doc content. 
            \label{fig:document_content}]{\includegraphics[width=1.0\linewidth, trim=152 92 152 92, clip]{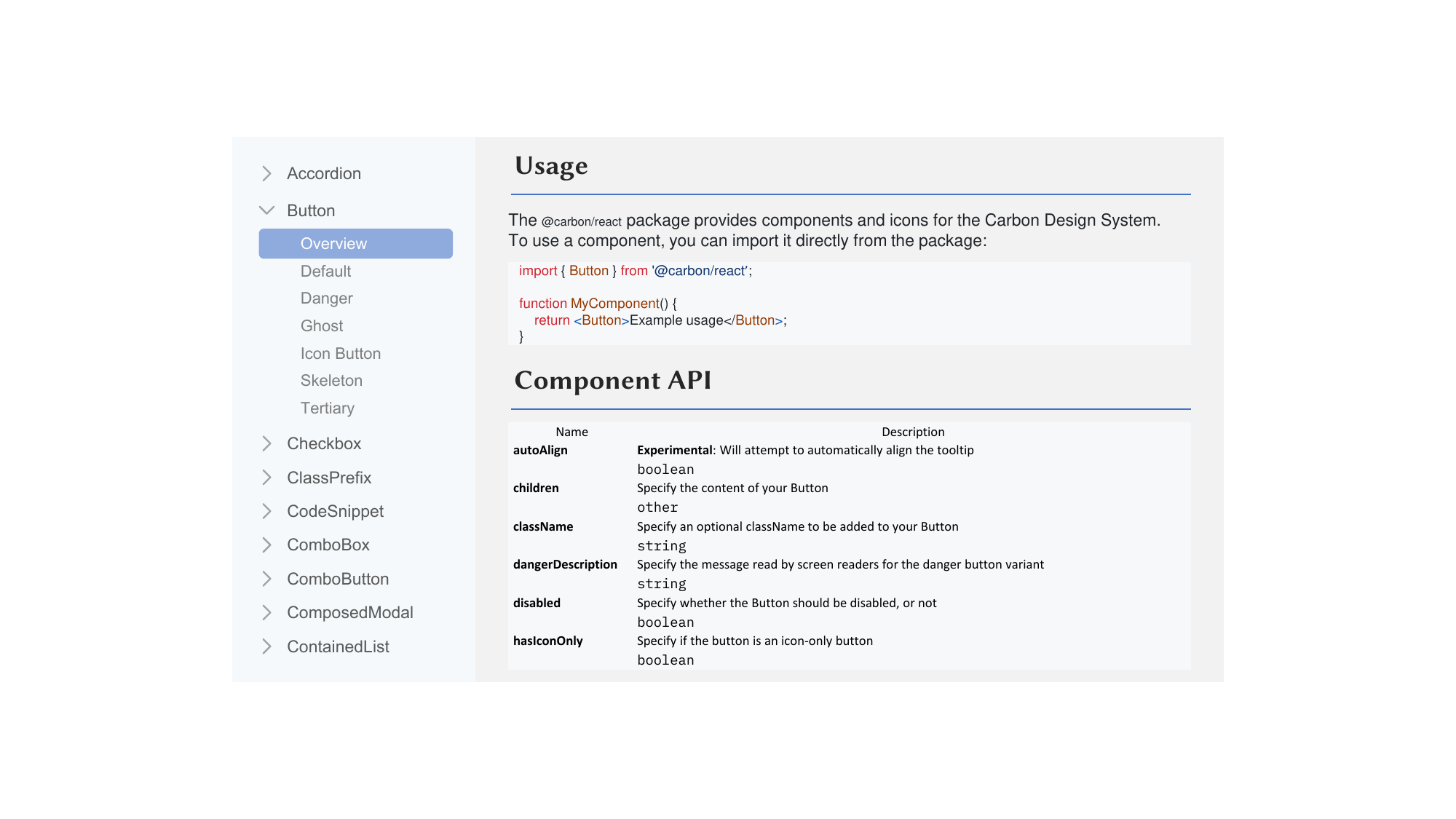}}
            \end{center}
    \vspace{-8px}
        \begin{center}
            \subfloat[Example of mined docs.\label{fig:doc_list_example}]{\includegraphics[width=1.0\linewidth, trim=225 64 225 64, clip]{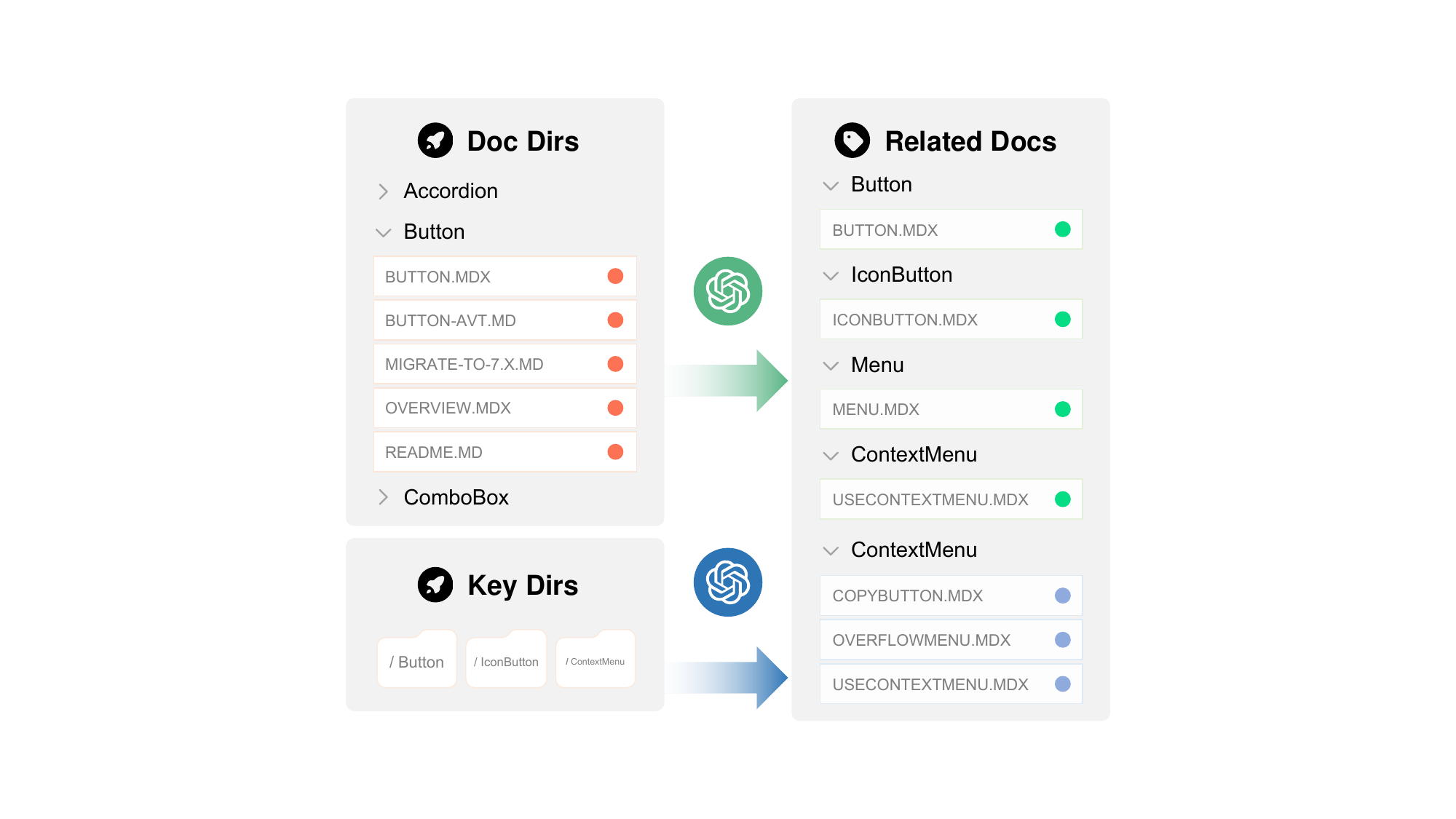}}
            \end{center}
    \vspace{-10px}
    \caption{Example of documents.} 
    \label{fig:knowledge_mining_example}
    \vspace{-10px}
\end{wrapfigure}

To facilitate effective reasoning, it is essential for the LLM to familiarize itself with the relevant project context such as usage patterns~\cite{Doc2Oracle}, so as to act more like a human developer. 
As shown in Figure~\ref{fig:document_content}, project documents (docs) contain important domain knowledge about what components are in the project and how to use them.
Hence, GUIRepair collects project-specific knowledge and incorporates it using the
knowledge injection paradigm~\cite{knowledgeInjection_Survey},
enabling the model to better comprehend the problem in subsequent reasoning tasks.
As illustrated in Figure~\ref{fig:guirepair_workflow}, GUIRepair takes both the issue report and project documentation as input. By processing the issue report, the model heuristically identifies and selects documents that are most relevant to the described scenario.
Specifically, we first prompt the chat model (e.g., GPT-4o~\cite{GPT4o}) to analyze the issue report and, based on its initial understanding, identify potentially relevant documents from the project's documentation directory. Due to the limited context window of LLMs, it is impractical to load and process all documents at once. Inspired by how human developers browse documentation, we instead provide only the directory structure, allowing the model to select the Top-N most relevant document filenames.
However, selecting documents solely based on filenames may miss important content, as the model cannot access the full text during this step. To address this, and inspired by retrieval-augmented generation (RAG) techniques~\cite{RAPGen}, we further use an embedding model (e.g., text-embedding-3~\cite{text-embedding-3-small}) to retrieve relevant content from the documentation repository. This retrieval is guided by the issue report and the chat model’s preliminary understanding. To reduce retrieval costs in large projects with hundreds or thousands of files, we limit the search to key directories identified in the previous chat model phase and return the Top-N documents with the highest similarity scores.
Ultimately, we obtain two sets of relevant documents: one selected by the chat model through directory browsing, and another retrieved via the embedding-based similarity search.
In Figure~\ref{fig:doc_list_example}, these two sets are merged to form a final collection of contextually relevant documents, which helps LLMs acquire project knowledge related to the issue scenario.

\subsubsection{Repro Generation}
Inspired by website generation~\cite{WebGen-Bench,DCGen} and bug reproduction~\cite{BugReproduce}, we design the reproduction (repro) generation step to enhance the model's understanding of visual problems by generating the corresponding code that reproduces the issue. To support this, we reorganize the model input by incorporating previously mined relevant documents as domain knowledge. This domain knowledge, which is closely related to the issue scenarios, is injected directly into the prompt as reference material~\cite{knowledgeInjection_Survey}.
\begin{wrapfigure}{r}{0.26\textwidth} 
    \vspace{-15px}
        \begin{center}
            \subfloat[The visual issue scenario of \textit{next-1509}. 
            \label{fig:image2code_image}]{\includegraphics[width=1.0\linewidth, trim=273 140 273 140, clip]{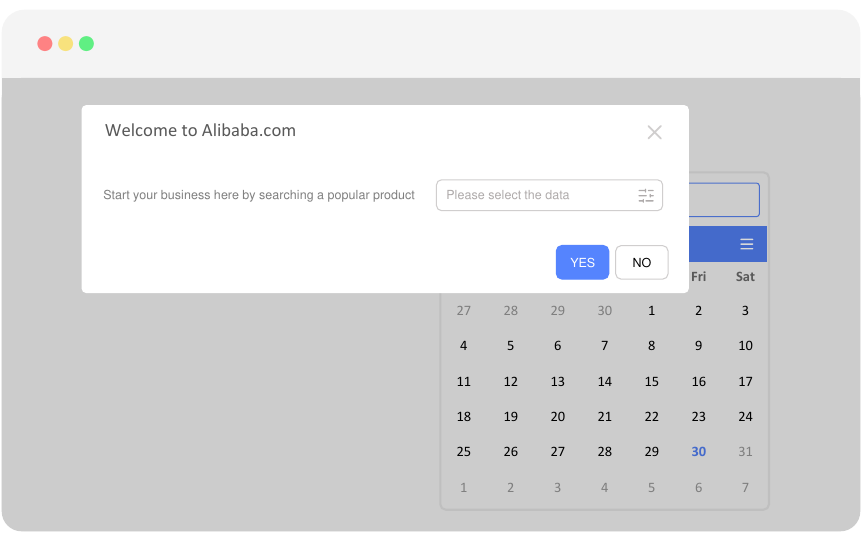}}
            \end{center}
    \vspace{-17px}
        \begin{center}
            \subfloat[The repro code in front-end project to replay the issue scenario of \textit{next-1509}.\label{fig:image2code_code}]{\includegraphics[width=1.0\linewidth, trim=273 140 273 140, clip]{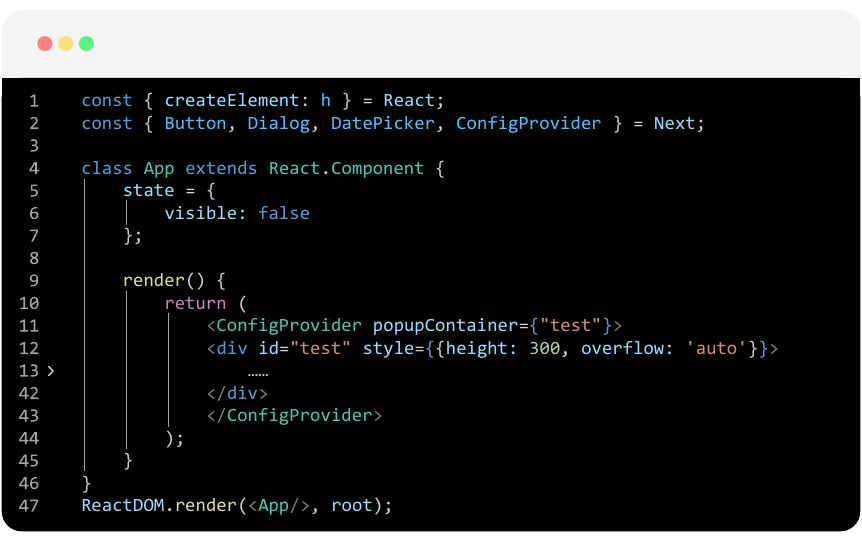}}
            \end{center}
    \vspace{-8px}
    \caption{Example of the reproduced code behind the front-end visual issue scenario (Image2Code).} 
    \label{fig:image2code_example}
    \vspace{-10px}
\end{wrapfigure}
As shown in Figure~\ref{fig:image2code_example}, we present the repro code behind a visual issue \textit{next-1509}~\cite{next-1509}. Figure~\ref{fig:image2code_image} shows the symptoms of the visual issue, where the calendar UI is incorrectly displayed below the dialog box, preventing users from directly using the calendar. The repro code in Figure~\ref{fig:image2code_code} reveals which code components were called in the visual problem (line 1-2) and how these code components were organized to trigger the problem scenario (line 4-47). These code details in Figure~\ref{fig:image2code_code} are crucial for understanding the visual problem and pinpointing the relevant code elements.
The goal of this step is to enable the model to learn project-specific knowledge in a manner similar to a human expert, thereby allowing it to reconstruct the code behind the visual issue. In many multimodal scenarios, reproducing an issue requires building a complete front-end project, typically composed of HTML, CSS, and JavaScript (JS) files~\cite{SWEBenchM}. Among these, the JS file contains the dynamic logic that reproduces the visual problem, while the HTML and CSS files generally contain static content that does not require frequent changes. Therefore, we instruct the model to generate only the JS code needed to reproduce the multimodal scenario. This output reveals the code components responsible for the visual problem and illustrates how they are structured to trigger the faulty behavior.
The generated code (as shown in Figure~\ref{fig:image2code_code}) is then embedded into the original issue report. The resulting enhanced issue report (i.e., Issue Report+), now includes the natural language description, the programming language reproduced code, and the visual representations of the issue. 
The enhanced issue report can be used in subsequent localization, generation, and validation processes to enhance fault understanding and implement fixes.
Indeed, an LLM may not always generate completely accurate reproduction code. Minor discrepancies may exist between the generated output and the original visual report, such as differences in button sizes or text highlight colors. However, this is tolerable, as our primary goal is to expose the underlying code behavior responsible for the visual problem. As long as the reproduced code sufficiently captures the essential behavior of the issue (e.g., API misuse), it serves its purpose in helping the model understand the multimodal scenario.
Furthermore, during the patch validation phase, we will re-render the issue scenario in a controlled local environment using the generated reproduction code. This eliminates noise from non-essential elements, allowing the LLM to focus more effectively on the core fault. As a result, minor detail discrepancies in the reproduced code have minimal impact on the repair process.


\subsection{Fault Localization}
\label{sec:fault_localization}
After obtaining the reproduced code and related documents for the visual problem, these project-specific domain knowledge will be used to enhance the fault localization process. As shown in Figure~\ref{fig:guirepair_workflow}, fault localization in GUIRepair is a two-stage process: \ding{184} file localization and \ding{185} hunk localization.

\subsubsection{File Localization}
In the file localization phase, GUIRepair leverages enhanced issue reports (with reproduced code) to understand the fault scenario and identify potentially related code files within the codebase. To achieve this, we inject project-specific documents into the chat model, enabling it to analyze the enhanced issue report and identify files likely to contain the bug. Since having the LLM read the entire codebase would exceed its context limit, we follow Agentless to extract the \textit{repository structure format}~\cite{Agentless}. Based on this structure, the model is prompted to return the paths of suspicious files after reviewing the overall layout of the codebase.
In addition, we use the embedding model to retrieve Top-N relevant code files from the codebase based on the enhanced issue report, helping to mitigate the chat model's limitation of not being able to view file content directly during the file localization phase.
Similar to the document retrieval strategy described in Section~\ref{sec:knowledge_mining}, the embedding model restricts its search to the directories identified by the chat model as likely containing the bug, which reduces the overhead of scanning the entire codebase by focusing only on files within key directories.
Finally, we merge the suspicious file paths identified by both the chat model and the embedding model to form the final set of suspicious files. 

Note that the final set may include a dozen or more files, which can exceed the model’s context limit during the hunk localization. To address this, we set a maximum number N of key bug files. If the file set exceeds this limit, we filter and retain only the Top-N most relevant files for further analysis.
In detail, we provide the LLM with the code content of each suspicious file and ask it to select Top-N key bug files. Inspired by the \textit{skeleton format}~\cite{Agentless}, we input a streamlined format of each file to avoid context limits. The original skeleton format retains class/function headers, variable declarations, and comments. GUIRepair extends this by preserving import statements to capture file-level dependencies and removing variable declarations to streamline the input. Based on this lightweight representation and its understanding of issue scenarios, the chat model identifies the Top-N key bug files.

\subsubsection{Hunk Localization}
In the hunk localization phase, we start with the list of key bug files and aim to identify the specific bug code hunks. We localize at the class or function level, rather than pinpointing exact code edit locations~\cite{Agentless}. This higher-level granularity simplifies the process while still capturing meaningful fault contexts.
Specifically, we provide the chat model with the full contents of all key bug files, allowing it to identify specific classes or functions that are likely responsible for the issue based on its understanding of the enhanced issue report. Unlike earlier stages where a compressed skeleton format is used, here we retain the complete file content. This is necessary because some bugs may appear outside of classes or functions, such as in configuration files or global code blocks, and the model needs full context to accurately navigate and localize the fault.
To alleviate context limitations, we compress only the content within classes and functions by defining a maximum hunk size of N lines. If a class or function exceeds this limit, we retain only its header, along with internal variable declarations and comments. The model then returns the names of all suspected bug-related classes or functions. For each identified element, we extract its complete class or function body as the localized hunk.
If the bug is located outside of a class or function such as in a configuration block or global scope, the model returns the relevant element name. In these cases, we define a context window (e.g., N lines) and extract a surrounding code segment that includes the bug element, which is then used as the localized hunk.

\subsection{Patch Generation}
\label{sec:patch_generation}
In the \ding{186} patch generation process, our goal is to generate candidate patches for the bug code snippets localized in the previous phase. Here, we follow the learning-based APR work~\cite{SequenceR,RewardRepair,SelfAPR,ITER,DLFix,DEAR,CoCoNut,CURE,KNOD,Recoder,Tare,NTR,Repilot,AlphaRepair,FitRepair,ChatRepair,SRepair,Contrastrepair,ThinkRepair,D4C,LLM4APR_Fan,LLM4APR_Huang_ASE,LLM4APR_Huang_TSE,LLM4APR_Jiang,LLM4APR_Wu,LLM4APR_Xia} to input class/function level bug code snippets and then ask the chat model to generate patches directly. 
For the format of the generated patches, we adopt the \textit{Search/Replace edit} strategy used in Agentless~\cite{Agentless}, due to its simplicity and effectiveness. A \textit{Search/Replace edit} consists of two key components: (1) a \textit{search} string, representing the original buggy code to be replaced, and (2) a \textit{replace} string, indicating the code snippet that should take its place. To integrate the patches into the project codebase, we search for occurrences of the specified \textit{search} snippet within the source files and replace them with the corresponding \textit{replace} snippet. This step enables seamless backfilling of the generated patches into the codebase.
Finally, we convert each patch into the unified git diff format using the \textit{git diff} command. These diff-formatted patches provide a clearer picture of the change behaviors and are used in the subsequent patch validation.

\begin{figure}[t]
        \begin{center}
            \subfloat[The bug/patch\_1/patch\_2 code behaviors in \textit{prism-1602}. 
            \label{fig:code2image_code}]{
            		\includegraphics[width=1.0\linewidth, trim=5 120 5 115, clip]{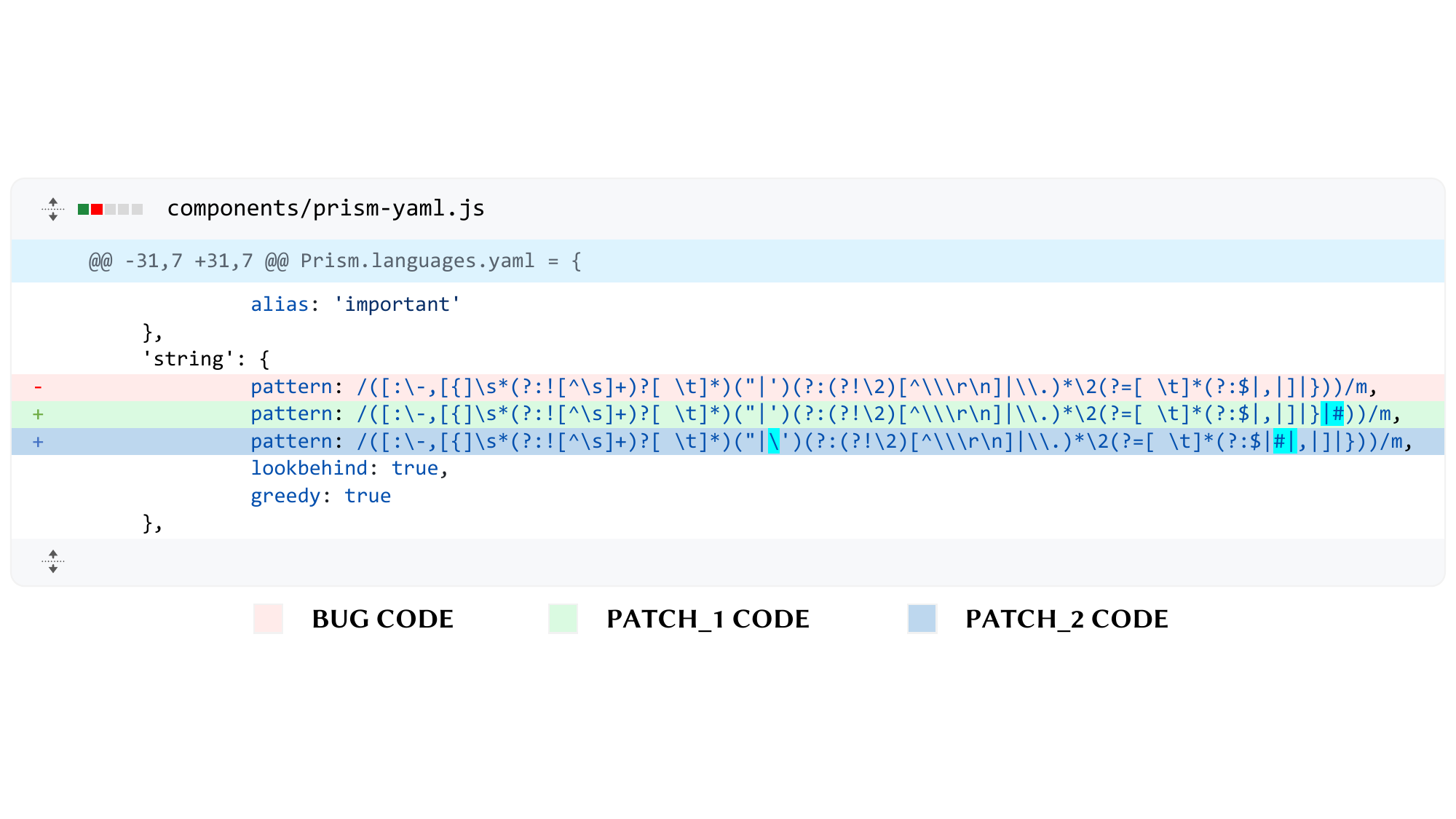}}
            \end{center}
        \begin{center}
            \subfloat[The visual images effects of bug/patch\_1/patch\_2 code in \textit{prism-1602}.\label{fig:code2image_image}]{
            		\includegraphics[width=1.0\linewidth, trim=58 202 58 205, clip]{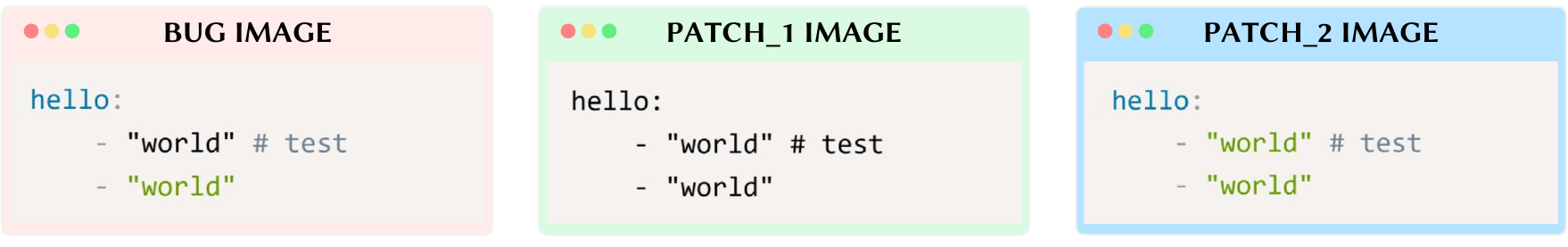}}
            \end{center}
    \vspace{-5px}
    \caption{Example of front-end visual image effects corresponding to bug/fix code behaviors (Code2Image).} 
    \label{fig:code2image_example}
    \vspace{-15px}
\end{figure}

\subsection{Patch Validation}
\label{sec:code2image}
In the LLM-driven APR workflow, LLMs typically sampled multiple times~\cite{LLM4APR_Huang_ASE,LLM4APR_Huang_TSE,LLM4APR_Wu,LLM4APR_Xia} to generate a diverse set of candidate patches. However, in the SWE task, a robust system is expected to achieve high accuracy using only its top-ranked patch, as measured by the Pass@1 evaluation metric. This requirement more accurately reflects the problem-solving capability of the AI system.
Therefore, the goal of the patch validation phase is to identify and select the most promising patches from the generated patch space. To facilitate this process, we introduce the Code2Image component, which assesses patches by capturing the visual representation of repair behavior. The validation process consists of two phases: \ding{187} GUI rendering, and \ding{188} patch selection.

\subsubsection{GUI Rendering}
In this phase, we aim to provide visual feedback for patch validation by capturing how the patch alters the third-party application's behavior as rendered in the GUI. To achieve this, we leverage the reproduced code generated by the LLM during the fault comprehension phase to construct a runnable front-end project. This project is then executed in a web browser to visually reflect the effect of the applied patch within the original bug scenario. As shown in Figure~\ref{fig:code2image_example}, different repair behaviors can produce different visual effects on the front end.
Although automatically replicating the execution environment would be ideal, it involves installing complex dependencies and adjusting JS package imports, which is an error-prone task for LLMs. As a practical compromise, we follow the procedure used in the SWE-bench M study~\cite{SWEBenchM} and perform this process manually.
First, we manually set up the local runtime environment for each JS project by following the repository’s developer or contributor guidelines and installing the necessary package manager and dependencies.
Second, we build front-end projects to replay the bugs using the reproduced code, manually importing the built JS package into the HTML file.  We then run the reproduced code in a local browser to capture a screenshot of the bug scenario. In some cases, minor manual modifications are needed to adjust package paths for correct execution.
Third, we apply each candidate patch from the patch space to the buggy repository. For each patched version, we rebuild the project to obtain a new JS package and run it in the browser. The resulting visual output is captured to represent the behavior of each patch. Finally, we obtain one bug scenario image and multiple patch scenario images. 

\subsubsection{Patch Selection}
In the patch selection phase, we evaluate the effectiveness of each patch using the previously captured visual information. 
Specifically, we first perform pixel-level comparisons between each patch scenario image and the bug scenario image. Patches that result in no visual change are considered ineffective and are filtered out. The remaining patch scenario images are then sequentially input to the LLM, which assesses whether the visual output reflects a successful fix based on its understanding of the issue report. Once the model identifies a patch as effective, the validation process terminates. The validated patch is then retained as the resolution for the issue.
To illustrate this process, consider the \textit{prism-1602} issue~\cite{prism-1602}, where YAML strings fail to highlight when a comment appears on the same line (i.e., ``world''). The buggy code (Figure~\ref{fig:code2image_code}) results in incorrect rendering in the front-end application (Figure~\ref{fig:code2image_image}). Our tool initially generates two candidate fixes: Patch\_1 and Patch\_2. Patch\_1 disables syntax highlighting entirely, while Patch\_2 correctly restores it. By observing these visual differences and associating with the previous understanding of issue symptoms, the LLM selects Patch\_2 as the effective fix. This example demonstrates how visual effects guide the model in choosing appropriate patches.

\section{Experiment Setup}

\subsection{Research Question}
\begin{itemize}[leftmargin=0.3cm]
    
    \item \textbf{RQ1: How well does the overall effectiveness of GUIRepair compare to state-of-the-art SWE autonomous systems?}
    This question evaluates GUIRepair's overall capabilities in solving multimodal issues. (Overall Effectiveness)

    \item \textbf{RQ2: How much does the design choice of GUIRepair contribute to
     the overall repair capability?}
    This question explores how Image2Code and Code2Image components affect the overall repair performance. (Ablation Study)

    \item \textbf{RQ3: Can GUIRepair generalize to additional multimodal task instances as well as other base models?}
    This question assesses GUIRepair's generalization ability on additional repositories and other models. (Generalizability Study)

\end{itemize}

\subsection{Benchmark}
We evaluate GUIRepair in SWE-bench M~\cite{SWEBenchM}, which is a benchmark for evaluating AI systems to fix bugs in visual, user-facing JavaScript software. It contains 619 task instances from 17 popular JavaScript libraries used for web interface design, diagramming, data visualization, syntax highlighting, and interactive mapping, each featuring images crucial to problem-solving. 
SWE-bench M is divided into two parts: 
Test split (SWE-bench M test) contains 517 instances. Development split (SWE-bench M dev) contains 102 instances. 
We use the standard test benchmark SWE-bench M test to evaluate the overall effectiveness, and the additional development set SWE-bench M dev for conducting the generalizability study.

\subsection{Baseline}
We compare GUIRepair against recent 13 SWE systems. These baseline tools represent the state-of-the-art performance on SWE-bench M. We directly select the top representative techniques on the official SWE-bench M leaderboard~\cite{Leaderboard} as of May 2025 (around the paper submission time) for comparison.
We include open-source as well as commercial or closed-source baselines, and the recent Computer Use Agent~\cite{ComputerUseAgent}.
We additionally provide a GPT-4o~\cite{GPT4o} based implementation for Agentless Lite~\cite{AgentlessLite} to serve as a complementary baseline, it is currently the best open-source system on SWE-bench M.

\subsection{Implementation}

We follow recent SWE studies~\cite{SWEBenchM,Agentless} to chose gpt-4o-2024-08-06 \cite{GPT4o} as the chat model and text-embedding-3-small \cite{text-embedding-3-small} as the embedding model for the implementation of GUIRepair. 
This is because GPT-4o solves more task instances at lower repair costs~\cite{SWEBenchM} and facilitates a fair comparison with baselines using a unified base model.
In terms of parameter settings, we follow the Agentless experience~\cite{Agentless,AgentlessLite}. 
By default, we query the chat model with greedy decoding (i.e., temperature = 0).
In the \ding{182} knowledge mining phase, we set the temperature for the chat model to 0, the sampling times to 1, and let the model return the Top-6 relevant documents. Then, we set the chunk size for the embedding model to 512, chunk overlap to 0, and select the Top-6 relevant documents retrieved by the model. Finally, we merge the results of the above two parts to get the final document list.
In the \ding{183} repro generation phase, we use the default chat model settings and one-shot prompting to generate the reproduced code.
In the \ding{184} file localization phase, we set the chat model's temperature of 1 and sampling time of 2 to obtain a diverse distribution, and ask the model to return all suspicious files. Besides, we keep the same settings for embedding model just like in the knowledge mining phase, and return Top-4 bug files. Similarity, we merge the results of the above two parts to get the final file list. In particular, we set the maximum number of candidate bug files to 4. If the final file list exceeds the maximum number of files, we ask the chat model (default settings) to read these candidates and filter out the Top-4 key files to alleviate the context limit.
In the \ding{185} hunk localization phase, we set the temperature of 1 and sampling time of 2 to cover potential bug code snippets/hunks to provide adequate contexts (or repair ingredients) for patch generation. In particular, for bug elements that are not inside the class/function, we set the context window to 500 to intercept the code hunk where the bug element is located.
In the \ding{186} patch generation phase, we use the greedy decoding strategy (default settings) to generate a patch and use the multi sampling strategy (temperature is 1 and sampling times is 39) to return 39 patches. Finally, we will get at most 40 patch candidates.
In the \ding{187} image capturing phase, we use \textit{fnm} to control different \textit{Node.JS} versions and use the \textit{npm/pnpm/yarn} as the JavaScript package manager, then we use Playwright~\cite{Playwright} to replay/capture the issue/patch scenario in browsers.
In the \ding{188} patch selection phase, we keep the default chat model settings and only submit the Top-1 valid patch to the SWE-bench evaluation platform.

\section{Evaluation}

\begin{table}[]
\caption{Results on SWE-bench M test.
{
\scriptsize
Note: 
\raisebox{-0.15em}{\includegraphics[height=0.9em]{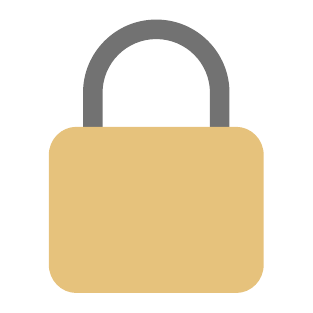}} refers closed-source commercial systems. 
\raisebox{-0.20em}{\includegraphics[height=1em]{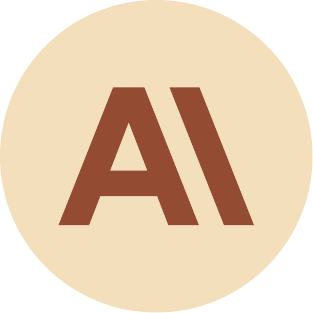}} refers claude-3-5-sonnet-20240620. 
\raisebox{-0.22em}{\includegraphics[height=1em]{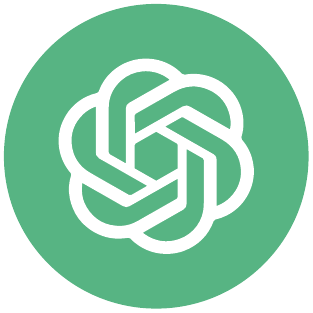}} refers gpt-4o-2024-08-06.} 
}
\label{tab:result_RQ1}
\resizebox{1.0\columnwidth}{!}{
\begin{tabular}{llccc}
\toprule
\rowcolor[HTML]{EFEFEF}
\multicolumn{1}{c}{\textbf{SWE System}} & \multicolumn{1}{c}{\textbf{Base Model}} & \textbf{Resolved} & \textbf{\%Resolved} & \textbf{\$Avg. Cost} \\
\midrule
\raisebox{-0.15em}{\includegraphics[height=1em]{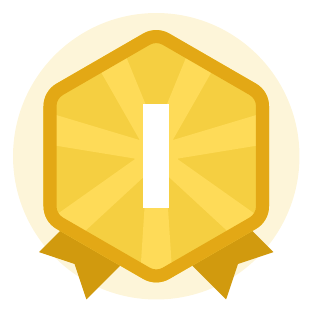}}
Globant Code Fixer Agent~\cite{Globant} \raisebox{-0.15em}{\includegraphics[height=0.9em]{Icon/lock}} & NA        & 153 & 29.59\% & -    \\
\rowcolor[HTML]{EFEFEF}
\raisebox{-0.20em}{\includegraphics[height=1em]{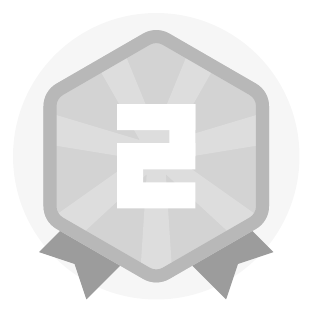}}
Zencoder~\cite{Zencoder} \raisebox{-0.15em}{\includegraphics[height=0.9em]{Icon/lock}}                 & NA        & 140 & 27.08\% & -    \\
\raisebox{-0.20em}{\includegraphics[height=1em]{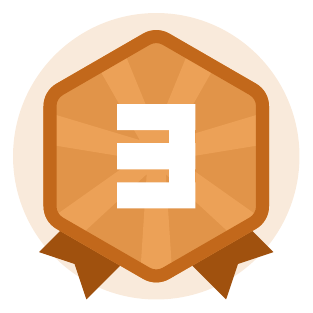}}
Agentless Lite~\cite{AgentlessLite}           & \raisebox{-0.20em}{\includegraphics[height=1em]{Icon/claude}} Claude3.5 & 131 & 25.34\% & \$0.25 \\
\rowcolor[HTML]{EFEFEF}
Agentless Lite~\cite{AgentlessLite}           & \raisebox{-0.22em}{\includegraphics[height=1em]{Icon/gpt}} GPT-4o     & 127 & 24.56\% & \$0.18    \\
Computer-Use Agents~\cite{ComputerUseAgent}      & \raisebox{-0.22em}{\includegraphics[height=1em]{Icon/gpt}} GPT-4o     & 104 & 20.12\% & -    \\
\rowcolor[HTML]{EFEFEF}
SWE-agent Multimodal~\cite{SWEBenchM}     & \raisebox{-0.22em}{\includegraphics[height=1em]{Icon/gpt}} GPT-4o     & 63  & 12.19\% & \$2.94 \\
SWE-agent~\cite{SWE-agent}                & \raisebox{-0.20em}{\includegraphics[height=1em]{Icon/claude}} Claude3.5 & 63  & 12.19\% & \$1.52 \\
\rowcolor[HTML]{EFEFEF}
SWE-agent JavaScript~\cite{SWEBenchM}     & \raisebox{-0.20em}{\includegraphics[height=1em]{Icon/claude}} Claude3.5 & 62  & 11.99\% & \$3.11 \\
SWE-agent~\cite{SWE-agent}                & \raisebox{-0.22em}{\includegraphics[height=1em]{Icon/gpt}} GPT-4o     & 62  & 11.99\% & \$2.07 \\
\rowcolor[HTML]{EFEFEF}
SWE-agent Multimodal~\cite{SWEBenchM}     & \raisebox{-0.20em}{\includegraphics[height=1em]{Icon/claude}} Claude3.5 & 59  & 11.41\% & \$3.11 \\
SWE-agent JavaScript~\cite{SWEBenchM}     & \raisebox{-0.22em}{\includegraphics[height=1em]{Icon/gpt}} GPT-4o     & 48  & 9.28\%  & \$0.99 \\
\rowcolor[HTML]{EFEFEF}
Agentless~\cite{Agentless}                & \raisebox{-0.20em}{\includegraphics[height=1em]{Icon/claude}} Claude3.5 & 32  & 6.19\%  & \$0.42 \\
RAG~\cite{SWEBenchM}                      & \raisebox{-0.22em}{\includegraphics[height=1em]{Icon/gpt}} GPT-4o     & 31  & 6.00\%  & \$0.17 \\
\rowcolor[HTML]{EFEFEF}
RAG~\cite{SWEBenchM}                      & \raisebox{-0.20em}{\includegraphics[height=1em]{Icon/claude}} Claude3.5 & 26  & 5.03\%  & \$0.15 \\
Agentless~\cite{Agentless}                & \raisebox{-0.22em}{\includegraphics[height=1em]{Icon/gpt}} GPT-4o     & 16  & 3.09\%  & \$0.38 \\
\rowcolor[HTML]{DFDCEF}
\raisebox{-0.23em}{\includegraphics[height=1em]{Icon/full}} GUIRepair                & \raisebox{-0.22em}{\includegraphics[height=1em]{Icon/gpt}} GPT-4o     & 157 & 30.37\% & \$0.29    \\
\bottomrule
\end{tabular}
}
\end{table}

\begin{table}[]
\caption{Resolved multimodol issues by repository.}
\label{tab:result_RQ1_repo}
\resizebox{1.0\columnwidth}{!}{
\begin{tabular}{lr|cccccc}
\toprule
\multicolumn{1}{c}{\multirow{2}{*}{\textbf{Repo}}} &
  \multicolumn{1}{c}{\multirow{2}{*}{\textbf{Num}}} &
  {GUIRepair} &
  {Globant~\cite{Globant}} &
  {Zencoder~\cite{Zencoder}} &
  {AgentlessL~\cite{AgentlessLite}} &
  {AgentlessL~\cite{AgentlessLite}} &
  {SWE-agentM~\cite{SWEBenchM}} \\
\multicolumn{1}{c}{}          & \multicolumn{1}{c}{} & \raisebox{-0.18em}{\includegraphics[height=1em]{Icon/gpt}}  & NA      & NA      & \raisebox{-0.18em}{\includegraphics[height=1em]{Icon/claude}} & \raisebox{-0.18em}{\includegraphics[height=1em]{Icon/gpt}}  & \raisebox{-0.18em}{\includegraphics[height=1em]{Icon/gpt}}  \\ 
\midrule
\rowcolor[HTML]{EFEFEF}
next           & 39                   & 9 (23.08\%)       & \cellcolor[HTML]{96D3A7}10 (25.64\%)     & 6 (15.38\%)      & 6 (15.38\%)        & 3 (7.69\%)      & 0 (0.00\%)      \\
\rowcolor[HTML]{FFFFFF}
bpmn-js               & 54                   & \cellcolor[HTML]{96D3A7}35 (64.81\%)     & 34 (62.96\%)     & 28 (51.85\%)     & 27 (50.00\%)       & 28 (51.85\%)     & 15 (27.78\%)     \\
\rowcolor[HTML]{EFEFEF}
carbon   & 134                  & 12 (8.96\%)     & 12 (8.96\%)     & \cellcolor[HTML]{96D3A7}15 (11.19\%)     & 9 (6.72\%)        & 9 (6.72\%)      & 2 (1.49\%)      \\
\rowcolor[HTML]{FFFFFF}
eslint                 & 11                   & \cellcolor[HTML]{96D3A7}6 (54.55\%)      & 2 (18.18\%)      & 2 (18.18\%)      & 2 (18.18\%)        & 1 (9.09\%)      & 0 (0.00\%)      \\
\rowcolor[HTML]{EFEFEF}
lighthouse       & 54                   & 3 (5.56\%)      & \cellcolor[HTML]{96D3A7}7 (12.96\%)      & \cellcolor[HTML]{96D3A7}7 (12.96\%)      & 4 (7.41\%)        & 3 (5.56\%)      & 3 (5.56\%)      \\
\rowcolor[HTML]{FFFFFF}
grommet               & 21                   & 1 (4.76\%)      & 1 (4.76\%)      & \cellcolor[HTML]{96D3A7}4 (19.05\%)      & 2 (9.52\%)        & 1 (4.76\%)      & 0 (0.00\%)      \\
\rowcolor[HTML]{EFEFEF}
highlight.js      & 39                   & \cellcolor[HTML]{96D3A7}4 (10.26\%)      & \cellcolor[HTML]{96D3A7}4 (10.26\%)      & \cellcolor[HTML]{96D3A7}4 (10.26\%)      & 3 (7.69\%)        & 1 (2.56\%)      & 1 (2.56\%)      \\
\rowcolor[HTML]{FFFFFF}
openlayers         & 79                   & \cellcolor[HTML]{96D3A7}76 (96.20\%)     & \cellcolor[HTML]{96D3A7}76 (96.20\%)     & 64 (81.01\%)     & 73 (92.41\%)       & 75 (94.94\%)     & 41 (51.90\%)     \\
\rowcolor[HTML]{EFEFEF}
prettier             & 13                   & \cellcolor[HTML]{96D3A7}2 (15.38\%)      & \cellcolor[HTML]{96D3A7}2 (15.38\%)      & 1 (7.69\%)      & 0 (0.00\%)        & 1 (7.69\%)      & 1 (7.69\%)      \\
\rowcolor[HTML]{FFFFFF}
prism                 & 38                   & \cellcolor[HTML]{96D3A7}7 (18.42\%)      & 4 (10.53\%)      & \cellcolor[HTML]{96D3A7}7 (18.42\%)      & 5 (13.16\%)        & 5 (13.16\%)      & 0 (0.00\%)      \\
\rowcolor[HTML]{EFEFEF}
quarto-cli         & 24                   & \cellcolor[HTML]{96D3A7}2 (8.33\%)      & 1 (4.17\%)      & \cellcolor[HTML]{96D3A7}2 (8.33\%)      & 0 (0.00\%)        & 0 (0.00\%)      & 0 (0.00\%)      \\
\rowcolor[HTML]{FFFFFF}
scratch-gui & 11                   & 0 (0.00\%)      & 0 (0.00\%)      & 0 (0.00\%)      & 0 (0.00\%)        & 0 (0.00\%)      & 0 (0.00\%)      \\ 
\midrule
\multicolumn{2}{c}{\textbf{Resolved}}          & 157     & 153     & 140     & 131       & 127     & 63      \\
\multicolumn{2}{c}{\textbf{\%Resolved}}        & 30.78\% & 30.00\% & 27.45\% & 25.69\%   & 24.90\% & 12.35\% \\
\bottomrule
\end{tabular}
}
\vspace{-10px}
\end{table}

\subsection{RQ1: Overall Effectiveness}

In the main experiment, we will investigate the overall effectiveness of GUIRepair on the SWE-bench M test. Table~\ref{tab:result_RQ1} presents the evaluation result of existing AI systems.
The effectiveness of GUIRepair in solving multimodal instances exceeds that of all SWE systems currently on the SWE-bench M leaderboard. Specifically, GUIRepair solves 4 more task instances than the best closed-source commercial system, Globant Code Fixer Agent~\cite{Globant}, and 17 more task instances than Zencoder~\cite{Zencoder}. This shows that GUIRepair's simple workflow is comparable to complex agent systems in terms of repair capability.
Besides, GUIRepair also solves 26 more task instances than the best open-source system Agentless Lite~\cite{AgentlessLite}.
Note that since closed-source systems do not release the base model they use, it is unfair to make a comparison directly with systems implemented with different models. Thus, if keeping the consistent base model, the GPT-4o based implementation of GUIRepair solves 30 more task instances than the best baseline Agentless Lite. Recall that GUIRepair adds design choices for multimodal scenarios to the agentless workflow. And the fact that our approach achieves better results than the general agentless system~\cite{Agentless,AgentlessLite}, which potentially suggests that our novel design further improves the ability of existing agentless systems to tackle visual problems.
Meanwhile, compared to agent systems~\cite{Globant,Zencoder,SWE-agent}, the agentless-based GUIRepair inherits the cost advantage of agentless approaches~\cite{Agentless,AgentlessLite}, which requires only \$0.29 on average to solve a multimodal instance.
Overall, GUIRepair is a success in terms of overall effectiveness. Not only does it remain relatively inexpensive in repair cost, but it rivals and surpasses state-of-the-art closed-source and open-source systems in terms of repair capability.


\begin{wrapfigure}{r}{0.26\textwidth}  
  \centering
  \vspace{-1px}
  {\includegraphics[width=1.0\linewidth, trim=286 90 280 94, clip]{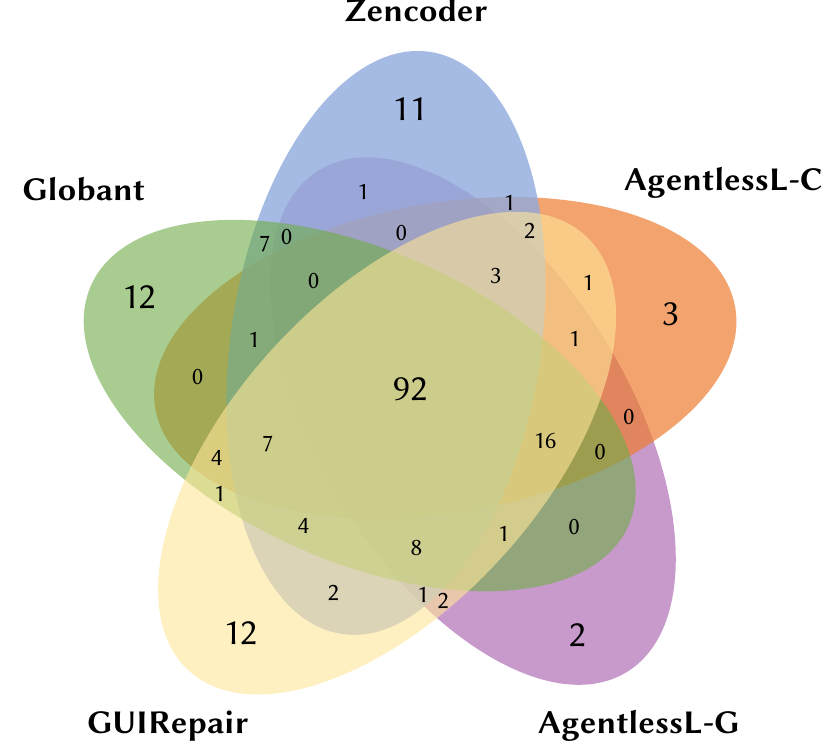}}
  \caption{Venn diagram for multimodal issue fixes. 
  {\scriptsize
  Note: AgentlessL-C/G refers Claude3.5/GPT4o-based Agentless Lite.}
  }
  \label{fig:venn_result}
  \vspace{-10px}
\end{wrapfigure}

Further, we will also explore the difference in repair capability between GUIRepair and existing systems in solving multimodal problems.
Here, we present the unique issues resolved by GUIRepair compared to existing Top-5 SWE systems in Figure~\ref{fig:venn_result}, and analyze the differences in the ability of these systems to resolve different repositories in Table~\ref{tab:result_RQ1_repo}.
As shown in Figure~\ref{fig:venn_result}, different SWE systems maintain unique fixes, suggesting that the different approaches differ in the characterization of the resolved issues. In particular, although GUIRepair is very close to the state-of-the-art closed-source commercial system Globant~\cite{Globant} (Globant Code Fixer Agent) in terms of repair capability (157 vs. 153), they both possess unique issue resloved (12 vs. 12), which suggests that GUIRepair solves task instances that are difficult to solve by the closed-source system, while the closed-source system at the same time compensates for GUIRepair's shortcomings in terms of repair capability.
Moreover, the overall distribution shows that GUIRepair complements rather than duplicates the capabilities of the other systems.
In addition, as shown in Table~\ref{tab:result_RQ1_repo}, different SWE systems have their strengths in solving the issues of different repositories. Specifically, we use the green background to highlight the best results from each repository. GUIRepair, Globant, and Zencoder achieved the best results in 7, 5, and 6 repositories. 
Overall, GUIRepair is orthogonal to existing systems in terms of repair capability, and it is competitive and complementary compared to top commercial systems.

\subsection{RQ2: Ablation Study}

In the ablation experiment, we aim to evaluate the contributions of GUIRepair’s key components (Image2Code and Code2Image) in enhancing the repair capability. Here, we designed four variants to reveal the effects of design choices:

\begin{itemize}[leftmargin=0.3cm]
    \item \raisebox{-0.20em}{\includegraphics[height=1em]{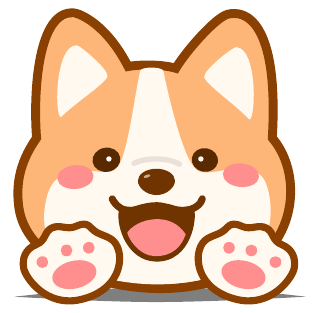}} \textbf{GUIRepair$_{base}$}: To provide a basic baseline to present the repair capabilities of the basic agentless approach in multimodal scenarios, we remove two key components (Section~\ref{sec:image2code} and~\ref{sec:code2image}) of the GUIRepair approach and keep only the core fault localization and patch generation modules (Section~\ref{sec:fault_localization} and \ref{sec:patch_generation}). This variant, called GUIRepair$_{base}$, receives multimodal issue reports and codebases directly as input and generates only one valid patch to be submitted.

    \item \raisebox{-0.20em}{\includegraphics[height=1em]{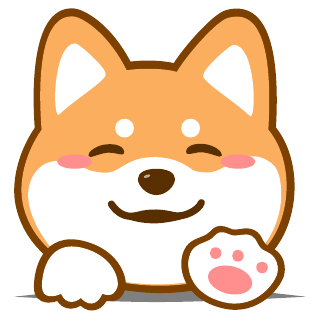}} \textbf{GUIRepair$_{I2C}$}: To present the improvement of the Image2Code component for the basic agentless workflow to solve multimodal scenarios, we add the Image2Code to the basic workflow. This variant, called GUIRepair$_{I2C}$, not only receives visual issue reports and codebases, but also analyzes the code behavior behind the visual information by mining project-specific knowledge to understand visual scenarios.

    \item \raisebox{-0.20em}{\includegraphics[height=1em]{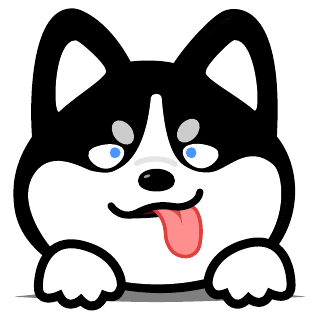}} \textbf{GUIRepair$_{C2I}$}: To indicate the enhancement of the Code2Image component for the basic agentless workflow to solve multimodal scenarios, we add the Code2Image to the basic workflow. This variant, called GUIRepair$_{C2I}$, also receives issue reports and codebases as input, in particular, it helps to select potentially correct patches from the patch space by capturing the visual effects of the repair behavior.

    \item \raisebox{-0.20em}{\includegraphics[height=1em]{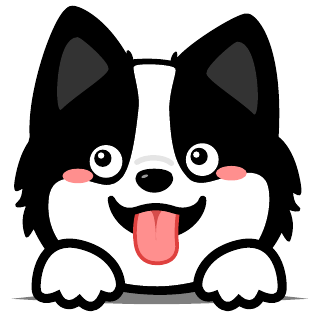}} \textbf{GUIRepair$_{full}$}: To explore the contribution of the two key components to overall effectiveness, we refer to the complete method implementation as GUIRepair$_{full}$. In fact, the two key components of GUIRepair, along with the basic agentless approach, form the complete workflow, and the process is an interdependent one, so that only the simultaneous use of the two key components can maximize the repair benefits.
\end{itemize}

As shown in Table~\ref{tab:result_RQ2}, we demonstrate the performance of different variants on SWE-bench M test. Next, we will analyze the impact of design choices on the repair capability.


\begin{table}[t]
\caption{Results of GUIRepair variants on SWE-bench M.}
\label{tab:result_RQ2}
\resizebox{1.0\columnwidth}{!}{
\begin{tabular}{l|cc|cc}
\toprule
\multicolumn{1}{c|}{\multirow{2}{*}{\textbf{Variants}}} & \multicolumn{2}{c|}{\textbf{Components}} & \multicolumn{2}{c}{\textbf{Result}} \\ 
\multicolumn{1}{c|}{} & Image2Code & Code2Image & Resolved  & Avg. \$Cost  \\ \midrule
\raisebox{-0.20em}{\includegraphics[height=1em]{Icon/base.pdf}} GUIRepair$_{base}$      & -         & -         & 136       & \$0.08     \\ \midrule
\raisebox{-0.20em}{\includegraphics[height=1em]{Icon/I2C.pdf}} GUIRepair$_{I2C}$       & \raisebox{-0.20em}{\includegraphics[height=1em]{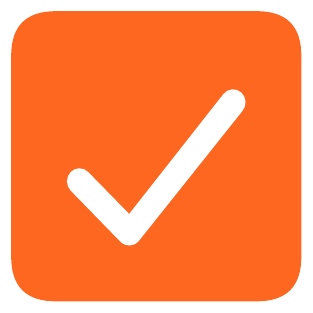}}        & -         & 146 (+10) & \$0.10     \\ \midrule
\raisebox{-0.20em}{\includegraphics[height=1em]{Icon/C2I.pdf}} GUIRepair$_{C2I}$       & -         & \raisebox{-0.20em}{\includegraphics[height=1em]{Icon/yes.pdf}}         & 148 (+12) & \$0.27     \\ \midrule
\raisebox{-0.20em}{\includegraphics[height=1em]{Icon/full.pdf}} GUIRepair$_{full}$   & \raisebox{-0.20em}{\includegraphics[height=1em]{Icon/yes.pdf}}         & \raisebox{-0.20em}{\includegraphics[height=1em]{Icon/yes.pdf}}         & 157 (+21) & \$0.29     \\  \bottomrule
\end{tabular}
}
\end{table}

\subsubsection{\raisebox{-0.20em}{\includegraphics[height=1em]{Icon/base.pdf}} 
Basic Capability of the Agentless Workflow}
As shown in Table~\ref{tab:result_RQ2}, GUIRepair$_{base}$ provides a reference baseline to indicate the performance achieved by the basic 
agentless workflow.
Specifically, GUIRepair$_{base}$ solves 136 instances, which is close to the performance of Agentless Lite (Table~\ref{tab:result_RQ1}). This is because both GUIRepair$_{base}$ and Agentless Lite follow the basic agentless workflow, and thus they are comparable in terms of repair capability. In particular, we note that GUIRepair$_{base}$ maintains a lower cost compared to existing agentless systems, Agentless Lite~\cite{AgentlessLite} and Agentless~\cite{Agentless,SWEBenchM}. This is due to the fact that our implementation further simplifies the fault localization process.
Compared to Agentless Lite, which locates faults by retrieving the codebase, GUIRepair$_{base}$ only retrieves the key file directories, thus reducing the file retrieval cost. Compared to Agentless, which uses three levels (file/element/edit-level) of fault location granularity to locate specific edit locations, GUIRepair$_{base}$ uses only two levels (file-level and hunk-level) of fault localization to locate hunk-level code snippets, thus further simplifying the complexity of the localization process. Benefiting from our simplified design, our implementation of the basic agentless approach has a lower cost advantage while maintaining good repair effectiveness.

\begin{figure}[t]
        \begin{center}
            \subfloat[The developer/GUIRepair$_{base}$/GUIRepair$_{I2C}$ patch of eslint-15243.\label{fig:case_study_i2c_1}]{
            		\includegraphics[width=1.0\linewidth, trim=65 85 65 85, clip]{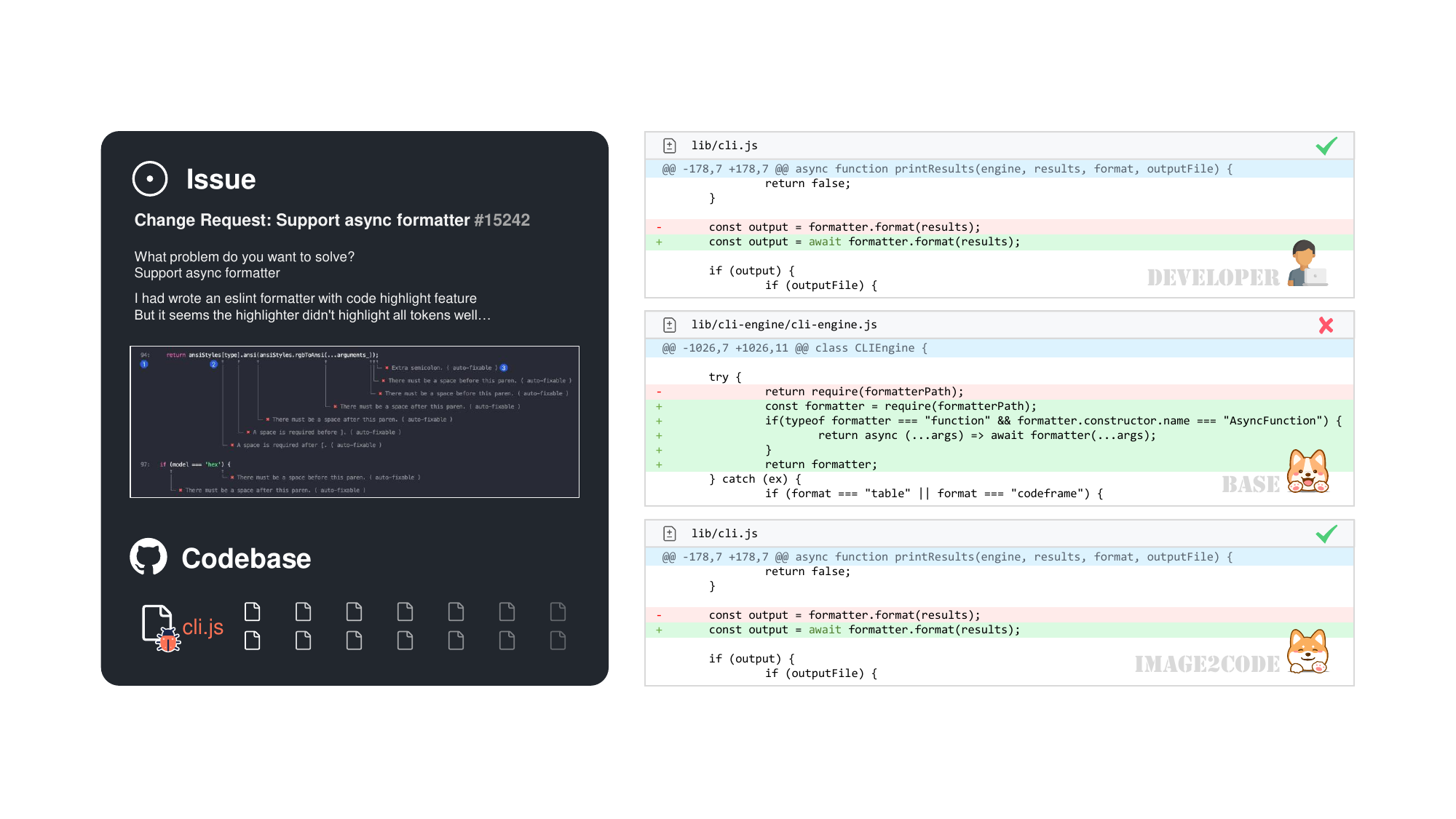}}
            \end{center}
        \begin{center}
            \subfloat[The reasoning trace of GUIRepair$_{base}$ to solve eslint-15243.\label{fig:case_study_i2c_2}]{
            		\includegraphics[width=1.0\linewidth, trim=65 210 65 210, clip]{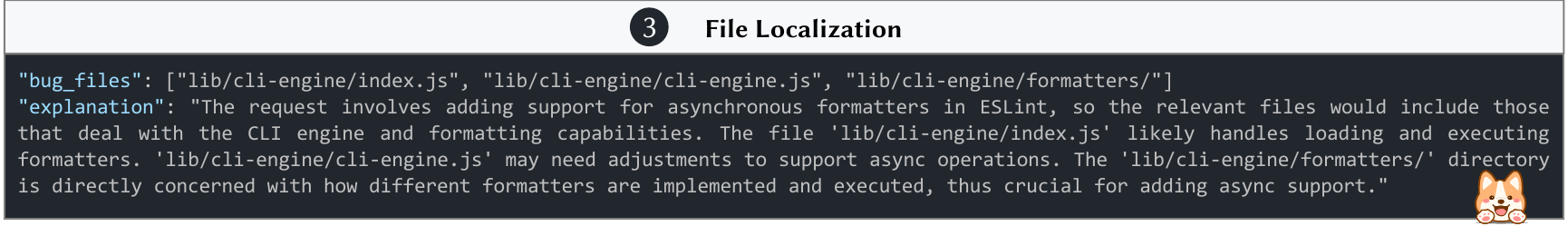}}
            \end{center}
        \begin{center}
            \subfloat[The reasoning trace of GUIRepair$_{I2C}$ to solve eslint-15243.\label{fig:case_study_i2c_3}]{
            		\includegraphics[width=1.0\linewidth, trim=65 115 65 115, clip]{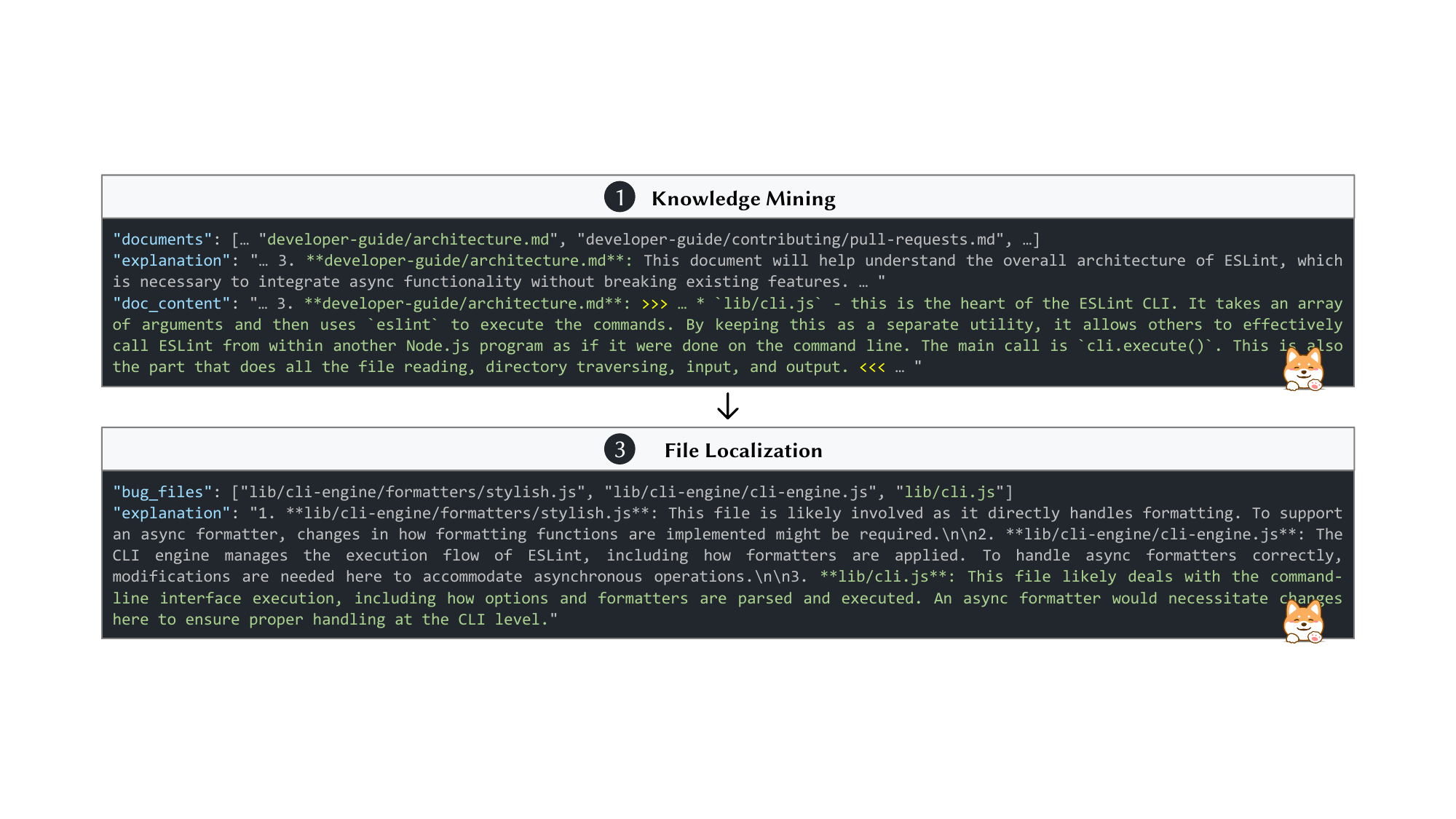}}
        \end{center}
    \vspace{-5px}
    \caption{The case study of eslint-15243.} 
    \label{fig:case_study_i2c}
    \vspace{-15px}
\end{figure}

\subsubsection{\raisebox{-0.20em}{\includegraphics[height=1em]{Icon/I2C.pdf}}
Improved Capability of the Image2Code Component}
As shown in Table~\ref{tab:result_RQ2}, GUIRepair$_{I2C}$ presents improvements of the Image2Code component to the basic repair workflow. Specifically, GUIRepair$_{I2C}$ solves 10 more instances than GUIRepair$_{base}$, achieving a 7.35\% improvement. Let's recall that one of the purposes of the Image2Code component is to enhance fault localization by mining the project-specific knowledge to understand the visual information.
To further clarify the contribution of the Image2Code, here we provide a case study to analyze how the Image2Code component works.
As shown in Figure~\ref{fig:case_study_i2c}, we demonstrate a instance \textit{eslint-15243}, which was successfully resolved by GUIRepair$_{I2C}$, but the GUIRepair$_{base}$ fix failed. In Figure~\ref{fig:case_study_i2c_1}, GUIRepair$_{I2C}$ accurately located the bug file \textit{cli.js} and generated a patch consistent with the developer patch. Yet GUIRepair$_{base}$ did not locate the correct bug file, which caused the fix to fail. Further, we try to analyze the reasoning trace behind these two variants to reveal why GUIRepair$_{I2C}$ was able to locate the bug file. As shown in Figure~\ref{fig:case_study_i2c_2}, GUIRepair$_{base}$ reads the issue report directly to locate suspicious files, however, since the model does not have project-specific knowledge, it struggles to accurately locate the relevant code files behind the visual issue. In contrast, in Figure~\ref{fig:case_study_i2c_3}, GUIRepair$_{I2C}$ first learns the project details in the knowledge mining phase, where it reads the key documents and learns the role of the \textit{cli.js}. Then, benefiting from the fact that the model has learned that project knowledge, when implementing fault localization, the model considers \textit{cli.js} as a suspicious file. In this way, GUIRepair$_{base}$ misses the opportunity to localize to the bug file due to its lack of project knowledge, while GUIRepair$_{I2C}$ compensates for this by using the Image2Code, which improves the overall effectiveness.

\subsubsection{\raisebox{-0.20em}{\includegraphics[height=1em]{Icon/C2I.pdf}}
Enhanced Capability of the Code2Image Component}
As shown in Table~\ref{tab:result_RQ2}, GUIRepair$_{C2I}$ presents enhancements of the Code2Image component to the basic agentless approach. 
Specifically, GUIRepair$_{C2I}$ solves 12 more multimodal task instances than GUIRepair$_{base}$, achieving a 8.82\% improvement. 
Such results indicate that the Code2Image component is critical for improving repair capabilities. It should be noted that although GUIRepair$_{C2I}$ achieves better results, it is almost three times more expensive to repair than GUIRepair$_{base}$. The reason for the increase in repair cost is mainly due to the multiple sampling performed by the patch generation in order to fully explore the patch space and the large amount of image information read when receiving visual feedback in patch selection phase. However, such a repair cost is still acceptable and it is still cheaper than many agent systems. Overall, capturing visual effects of repair behaviors to provide fix feedback is an effective validation scheme for visual issues.

\begin{figure}[t]
        \begin{center}
            \subfloat[The issue report of next-4182. 
            {\tiny Note: left is the textual description, right is the issue image.}\label{fig:case_study_full_1}]{
            		\includegraphics[width=1.0\linewidth, trim=65 103 65 105, clip]{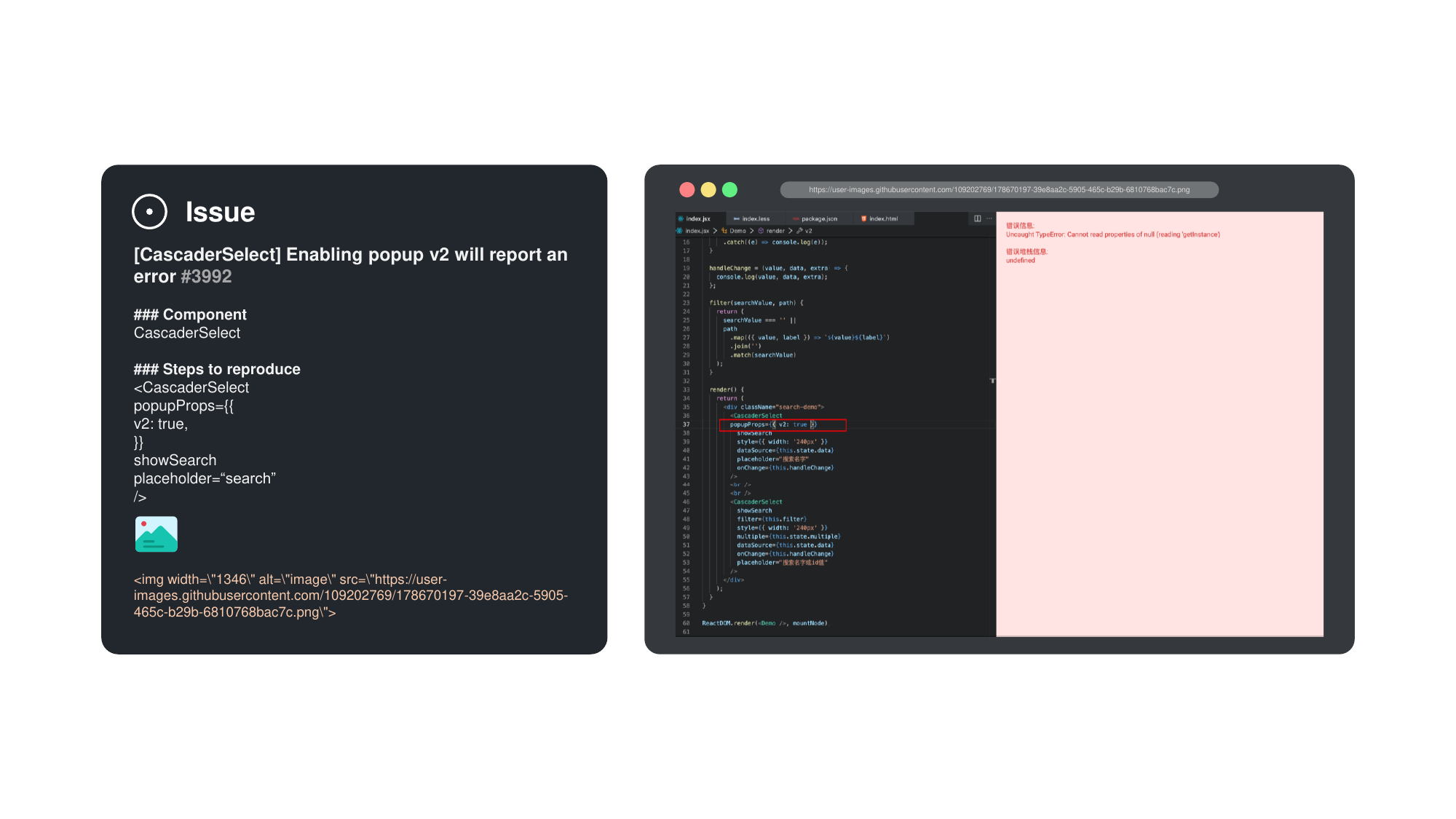}}
            \end{center}
        \begin{center}
            \subfloat[The reasoning trace of GUIRepair$_{full}$ to solve next-4182.\label{fig:case_study_full_2}]{
            		\includegraphics[width=1.0\linewidth, trim=65 165 65 165, clip]{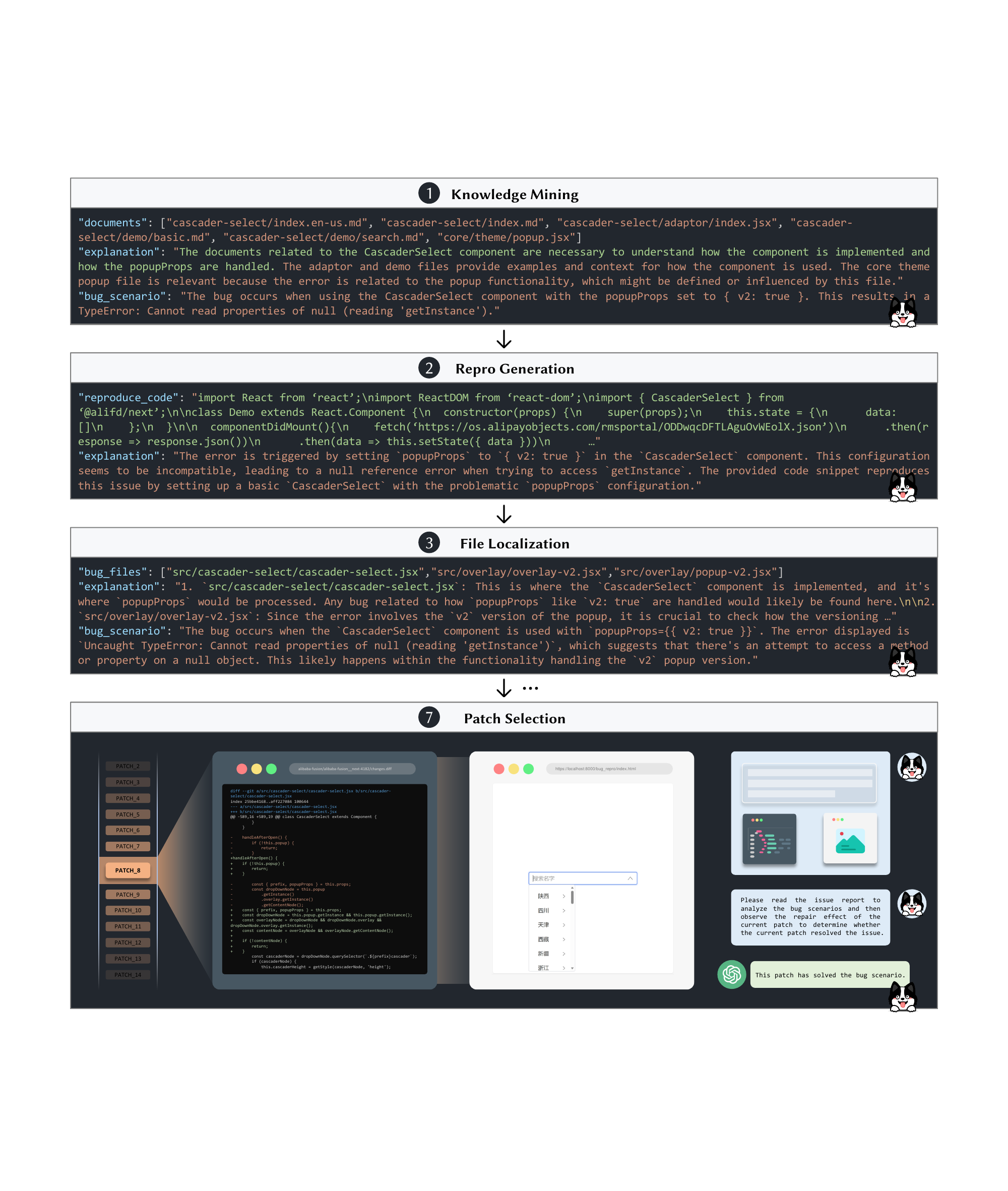}}
            \end{center}
    \vspace{-5px}
    \caption{The case study of next-4182.} 
    \label{fig:case_study_full}
    \vspace{-15px}
\end{figure}

\subsubsection{\raisebox{-0.20em}{\includegraphics[height=1em]{Icon/full.pdf}}
Full Capability of the Complete Pipeline}
We have presented the unique contributions of two key components. However, using any single component alone cannot fully unlock the complete performance of GUIRepair. On the one hand, Image2Code-only GUIRepair$_{I2C}$ only helps to understand visual scenarios by mining project-specific knowledge for improving fault localization, ignoring a crucial role (i.e., guided patch validation) of reasoning the code behavior behind the image information. On the other hand, Code2Image-only GUIRepair$_{C2I}$ only works on instances with reproduced code, which means it cannot solve most of the visual problems (only around 17\% of instances contain reproduced link).
As shown in Table~\ref{tab:result_RQ2}, the full implementation, GUIRepair$_{full}$, solves 157 instances, which is better than any other variant. Such results suggest that implementing the full workflow simultaneously is the only way to maximize the fixing benefits. 
To clarify more clearly how the two components work together,
we show an instance \textit{next-4182} that can only be resolved by GUIRepair$_{full}$. As shown in Figure~\ref{fig:case_study_full_1}, the issue report does not provide the complete repro code, which results in GUIRepair$_{C2I}$ unable to implement patch validation. Furthermore, even though GUIRepair$_{I2C}$ is able to generate the repro code, it is not equipped with the Code2Image module, resulting in the inability to fully utilize the repro generation. Therefore, these variants cannot solve this case. 
Here, we show the reasoning trace of GUIRepair$_{full}$ in Figure~\ref{fig:case_study_full_2}. Using Image2Code, the model successfully generates repro code (\ding{183}~Repro Generation) for the bug scenario after learning project-specific knowledge (\ding{182}~Knowledge Mining) and helps the model locate the correct bug file \textit{cascader-select.jsx} (\ding{184}~File Localization). 
Then, the Code2Image component uses the previously generated repro code to capture the visual effects presented by the fixing behavior. When the iteration validates to Patch\_8 (\ding{188}~Patch Selection), the model observes that the bug component renders properly on the web page and considers the patch resolved this issue. In this case, the reproduced code is a key factor in helping to solve the problem by tightly linking the Image2Code and Code2Image components to better understand the bug and validate the patch. Since most visual instances lack reproduction links, the full repair potential can only be unlocked by using both key components together. Overall, GUIRepair requires not only the Image2Code to understand the visual problem and generate the reproduced code, but also the Code2Image to use the reproduced code to capture the visual effect of the fixing behavior, which is an inextricably intertwined process.

\begin{table}[t]
\caption{Results of GUIRepair on additional repositories. 
{\scriptsize Note: all systems listed here use GPT-4o as the base model.}
}
\label{tab:result_RQ3_repo}
\resizebox{1.0\columnwidth}{!}{
\begin{tabular}{lr|ccccc}
\toprule
\rowcolor[HTML]{FFFFFF}
\multicolumn{1}{c}{\textbf{Repo}} &
  \multicolumn{1}{c|}{\textbf{Num}} &
  \multicolumn{1}{l}{\textbf{GUIRepair$_{full}$}} &
  \multicolumn{1}{l}{\textbf{GUIRepair$_{base}$}} &
  \multicolumn{1}{l}{\textbf{SWE-agentM}} &
  \multicolumn{1}{l}{\textbf{AgentlessJS}} &
  \multicolumn{1}{l}{\textbf{RAG}} \\ \midrule
\rowcolor[HTML]{EFEFEF}
wp-calypso               & 37               & 0 & 0 & 0  & 0 & 1  \\
\rowcolor[HTML]{FFFFFF}
Chart.js                 & 24               & 6 & 5 & 5  & 0 & 3  \\
\rowcolor[HTML]{EFEFEF}
react-pdf                & 11               & 0 & 0 & 0  & 0 & 0  \\
\rowcolor[HTML]{FFFFFF}
marked                   & 14               & 1 & 1 & 1  & 0 & 3  \\
\rowcolor[HTML]{EFEFEF}
p5.js                    & 16               & 7 & 4 & 4  & 1 & 4  \\
\midrule
\rowcolor[HTML]{DFDCEF}
\multicolumn{2}{c|}{\textbf{Total Resolved}} & \textbf{14} & 10 & 10 & 1 & 11 \\
\bottomrule
\end{tabular}
}
\end{table}

\begin{table}[]
\caption{Results of GUIRepair with different base models.}
\label{tab:result_RQ3_model}
\resizebox{1.0\columnwidth}{!}{
\begin{tabular}{lr|cc|cc|cc}
\toprule
\multicolumn{1}{c}{\multirow{2}{*}{\textbf{Repo}}} & 
\multicolumn{1}{c|}{\multirow{2}{*}{\textbf{Num}}}  & 
\multicolumn{2}{c|}{\textbf{GPT-4o}}                & 
\multicolumn{2}{c|}{\textbf{GPT-4.1}}               & 
\multicolumn{2}{c}{\textbf{o4-mini}} \\ 
\multicolumn{1}{c}{}                                & 
\multicolumn{1}{c|}{}                                & 
\raisebox{-0.20em}{\includegraphics[height=1em]{Icon/base}}$_{base}$   & 
\raisebox{-0.20em}{\includegraphics[height=1em]{Icon/full}}$_{full}$   & 
\raisebox{-0.20em}{\includegraphics[height=1em]{Icon/base}}$_{base}$   & 
\raisebox{-0.20em}{\includegraphics[height=1em]{Icon/full}}$_{full}$   & 
\raisebox{-0.20em}{\includegraphics[height=1em]{Icon/base}}$_{base}$   & 
\raisebox{-0.20em}{\includegraphics[height=1em]{Icon/full}}$_{full}$ \\
\midrule
\rowcolor[HTML]{EFEFEF}
next         & 39  & 6  & 9  & 5  & 8  & 6  & 10  \\
\rowcolor[HTML]{FFFFFF}
bpmn-js      & 54  & 35 & 35 & 32 & 33 & 37 & 37  \\
\rowcolor[HTML]{EFEFEF}
carbon       & 134 & 7  & 12 & 15 & 16 & 14 & 19  \\
\rowcolor[HTML]{FFFFFF}
eslint       & 11  & 2  & 6  & 1  & 5  & 5  & 6  \\
\rowcolor[HTML]{EFEFEF}
lighthouse   & 54  & 3  & 3  & 3  & 4  & 6  & 6  \\
\rowcolor[HTML]{FFFFFF}
grommet      & 21  & 1  & 1  & 2  & 2  & 2  & 4  \\
\rowcolor[HTML]{EFEFEF}
highlight.js & 39  & 4  & 4  & 3  & 3  & 3  & 3  \\
\rowcolor[HTML]{FFFFFF}
openlayers   & 79  & 76 & 76 & 76 & 76 & 77 & 77  \\
\rowcolor[HTML]{EFEFEF}
prettier     & 13  & 0  & 2  & 1  & 2  & 4  & 4  \\
\rowcolor[HTML]{FFFFFF}
prism        & 38  & 1  & 7  & 7  & 9  & 4  & 6  \\
\rowcolor[HTML]{EFEFEF}
quarto-cli   & 24  & 1  & 2  & 3  & 3  & 2  & 3   \\
\rowcolor[HTML]{FFFFFF}
scratch-gui  & 11  & 0  & 0  & 0  & 0  & 0  & 0  \\ 
\midrule
\rowcolor[HTML]{DFDCEF}
\multicolumn{2}{c|}{\textbf{Total Resolved}} & 136  & \textbf{157}  & 148  & \textbf{161}  & 160  & \textbf{175}  \\
\rowcolor[HTML]{DFDCEF}
\multicolumn{2}{c|}{\textbf{Avg. \$Cost}}    & \$0.08 & \$0.29 & \$0.07  & \$0.25 & \$0.08 & \$0.36  \\
\bottomrule
\end{tabular}
}
\end{table}

\subsection{RQ3: Generalizability Study}
In the generalizability experiment, we focus on the performance of GUIRepair on additional projects and base models.
We follow the recent study~\cite{SWEBenchM} testing the performance of GUIRepair on SWE-bench M dev to investigate whether our approach can be generalized to additional projects.
We choose GPT-4.1~\cite{GPT-4.1} (flagship GPT model for complex tasks) and o4-mini~\cite{o4-mini} (faster, more affordable reasoning model) to analyze whether our method can work effectively on other base models.
As shown in Table~\ref{tab:result_RQ3_repo}, GUIRepair achieves the best results on SWE-bench M dev. In particular, GUIRepair$_{full}$ solves 4 more instances than GUIRepair$_{base}$, showing that our approach is equally effective in other projects.
In addition, we show the performance of GUIRepair on different models in Table~\ref{tab:result_RQ3_model}.
Specifically, GUIRepair$_{full}$ resolves 21, 13, and 15 more instances than GUIRepair$_{base}$ when using GPT-4o, GPT-4.1, and o4-mini, respectively. Notably, with the reasoning model o4-mini, GUIRepair outperforms the best commercial system, Globant, by 22 instances (175 vs. 153). Such results show that our approach is not only effective across different models, but also significantly benefits from enhanced reasoning capabilities~\cite{Resoning4APR}, surpassing SOTA commercial solutions.
Overall, these findings highlight the generalizability of GUIRepair across diverse projects and model backbones.

\section{Threats to Validity}

\noindent
\textbf{Data Leakage.}
To mitigate the threat of data leakage, we keep the same model as baselines. In the main experiment (Table~\ref{tab:result_RQ1}), GUIRepair based on GPT-4o still performs better than baselines using the same model. 
This suggests that even under the same threat of data leakage, our approach still outperforms other schemes. 
Moreover, our ablation study (Table~\ref{tab:result_RQ2}) and generalizability study (Table~\ref{tab:result_RQ3_model}) show that the improvement does not only stem from the powerful base model, but is further enhanced by the design choices of GUIRepair.

\noindent
\textbf{Repo Configuration.}
We manually install dependencies and deploy runtime environments for each project, which is an inefficient process. Fortunately, SWE-bench M just released docker images. Besides, we see the potential for autonomous systems to replace humans in deploying and running code repository~\cite{ExecutionAgent}, and we plan to use these automated techniques in the future to replace manual/hard-encoding practices.

\section{Related Work}


\noindent
\textbf{LLM for APR.}
Researchers have proposed many LLM-based APR tools to solve software engineering problems. Early researchers fixed real-world software defects by fine-tuning~\cite{LLM4APR_Huang_ASE,LLM4APR_Huang_TSE,LLM4APR_Jiang,LLM4APR_Wu,NTR,RAPGen,FitRepair,MoRepair,RepairLlama} or prompting~\cite{LLM4APR_Fan,LLM4APR_Xia,AlphaRepair,GAMMA,ChatRepair,ThinkRepair,RepairAgent,FixAgent,Adversarial,WILLIAMT,DRCodePilot,D4C,CREF,PredicateFix,TypeFix,SRepair} LLMs and showed promising results. Recently, researchers~\cite{SWE-agent,Agentless,AutoCodeRover,SpecRover,LingmaSWE,LingmaAgent,Openhands,Moatless,Marscode,APRAgent_Study} have begun to focus on github issue resolution~\cite{SWEBench,SWEBenchM}, and proposed many agent/agentless-based autonomous systems to solve complex repo-level issue scenarios.
For example, SWE-agent~\cite{SWE-agent} solves SWE problems by designing Agent Computer Interface to interact with the environment, SpecRover~\cite{SpecRover} guides the LLM fixing process by specifications inference, and Agentless~\cite{LingmaAgent} employs a simplistic pipeline of localization, repair, and validation to solve software development problems, etc. 
However, most existing SWE systems are not designed for multimodal scenarios, which makes them difficult to solve visual problems~\cite{SWEBenchM}. 
In addition, although some studies also focus on visual-related issues, they are often designed only for specific bug types and are difficult to generalize to generic visual problem domains. 
For example, DesignRepair~\cite{Designrepair} only focuses on UI design-related quality issues, Iris~\cite{Iris} is designed for color-related accessibility issues. 
As shown in Figure~\ref{fig:instances_example}, multimodal task instances contain various issue scenarios, leading to the need for more generalized solutions to be applied in the vision software domain.
GUIRepair builds on simple agentless workflows~\cite{Agentless,AgentlessLite} and enhances the ability to solve multimodal issues by leveraging cross-modal reasoning to understand and capture visual information.

\noindent
\textbf{LLM for GUI.}
GUI-related problems directly affect the user experience. A key area of research is GUI testing, especially researchers have used LLM-driven approaches to implement GUI testing~\cite{LLM4GUI_Survey,LLM4GUIAPP_Survey}.
For example, researchers explored applications by enabling LLMs to understand multimodal scenarios and simulate user actions to test and reproduce GUI-related issues~\cite{GPTDroid,AdbGPT,SoapTesting}.
Especially, recent work has investigated reproducing bugs from bug reports~\cite{Report2Test_Fazz,Report2Test_Zhao} or bug videos~\cite{Video2Test_Ber,Video2Test_Hav}.
And this key reproduction information has successfully helped to implement soap opera tests generation~\cite{SoapTesting_Report,SoapOperaTg}.
Inspired by GUI testing work, GUIRepair implements fault comprehension and patch validation by generating reproduced code and replaying visual scenarios to migrate the experience of GUI testing to APR.
\section{Conclusion}
In this work, we propose GUIRepair, a cross-modal reasoning agentless approach that designs Image2Code and Code2Image components for understanding and capturing visual information to enhance the capabilities of autonomous systems for multimodal SWE problems.
Our evaluation on the SWE-bench M demonstrates that GUIRepair can achieve the highest performance compared to other open/closed-source systems.


\bibliographystyle{IEEEtran}
\bibliography{sample-base}

\begin{thebibliography}{10}
\providecommand{\url}[1]{#1}
\csname url@samestyle\endcsname
\providecommand{\newblock}{\relax}
\providecommand{\bibinfo}[2]{#2}
\providecommand{\BIBentrySTDinterwordspacing}{\spaceskip=0pt\relax}
\providecommand{\BIBentryALTinterwordstretchfactor}{4}
\providecommand{\BIBentryALTinterwordspacing}{\spaceskip=\fontdimen2\font plus
\BIBentryALTinterwordstretchfactor\fontdimen3\font minus \fontdimen4\font\relax}
\providecommand{\BIBforeignlanguage}[2]{{%
\expandafter\ifx\csname l@#1\endcsname\relax
\typeout{** WARNING: IEEEtran.bst: No hyphenation pattern has been}%
\typeout{** loaded for the language `#1'. Using the pattern for}%
\typeout{** the default language instead.}%
\else
\language=\csname l@#1\endcsname
\fi
#2}}
\providecommand{\BIBdecl}{\relax}
\BIBdecl

\bibitem{LLM4APR_Report}
C.~Le~Goues, M.~Pradel, A.~Roychoudhury, and S.~H. Tan, ``{Automated Programming and Program Repair (Dagstuhl Seminar 24431)},'' \emph{Dagstuhl Reports}, vol.~14, no.~10, pp. 39--57, 2025.

\bibitem{APR_CACM}
C.~Le~Goues, M.~Pradel, and A.~Roychoudhury, ``Automated program repair,'' \emph{Communications of the ACM (CACM)}, vol.~62, no.~12, pp. 56--65, 2019.

\bibitem{APR_Survey_CSUR}
K.~Huang, Z.~Xu, S.~Yang, H.~Sun, X.~Li, Z.~Yan, and Y.~Zhang, ``Evolving paradigms in automated program repair: Taxonomy, challenges, and opportunities,'' \emph{ACM Computing Surveys (CSUR)}, vol.~57, no.~2, pp. 1--43, 2024.

\bibitem{APR_Survey_TOSEM}
Q.~Zhang, C.~Fang, Y.~Ma, W.~Sun, and Z.~Chen, ``A survey of learning-based automated program repair,'' \emph{ACM Transactions on Software Engineering and Methodology (TOSEM)}, vol.~33, no.~2, pp. 1--69, 2023.

\bibitem{APR_Survey_Arxiv}
Q.~Zhang, C.~Fang, Y.~Xie, Y.~Ma, W.~Sun, Y.~Yang, and Z.~Chen, ``A systematic literature review on large language models for automated program repair,'' \emph{arXiv preprint arXiv:2405.01466}, 2024.

\bibitem{APR_Survey_Renzullo}
J.~Renzullo, P.~Reiter, W.~Weimer, and S.~Forrest, ``Automated program repair: Emerging trends pose and expose problems for benchmarks,'' \emph{ACM Computing Surveys (CSUR)}, vol.~57, no.~8, pp. 1--18, 2025.

\bibitem{APR_industry}
H.~Eladawy, C.~Le~Goues, and Y.~Brun, ``Automated program repair, what is it good for? not absolutely nothing!'' in \emph{46th International Conference on Software Engineering (ICSE)}, 2024, pp. 1--13.

\bibitem{LLM4SE_survey}
Q.~Zhang, C.~Fang, Y.~Xie, Y.~Zhang, Y.~Yang, W.~Sun, S.~Yu, and Z.~Chen, ``A survey on large language models for software engineering,'' \emph{arXiv preprint arXiv:2312.15223}, 2023.

\bibitem{LLM4SE_Survey_Arxiv}
J.~Liu, K.~Wang, Y.~Chen, X.~Peng, Z.~Chen, L.~Zhang, and Y.~Lou, ``Large language model-based agents for software engineering: A survey,'' \emph{arXiv preprint arXiv:2409.02977}, 2024.

\bibitem{RepairAgent}
I.~Bouzenia, P.~Devanbu, and M.~Pradel, ``Repairagent: An autonomous, llm-based agent for program repair,'' in \emph{47th International Conference on Software Engineering (ICSE)}, 2025, pp. 694--694.

\bibitem{FixAgent}
C.~Lee, C.~S. Xia, L.~Yang, J.-t. Huang, Z.~Zhu, L.~Zhang, and M.~R. Lyu, ``A unified debugging approach via llm-based multi-agent synergy,'' \emph{arXiv preprint arXiv:2404.17153}, 2024.

\bibitem{ChatRepair}
C.~S. Xia and L.~Zhang, ``Automated program repair via conversation: Fixing 162 out of 337 bugs for \$0.42 each using chatgpt,'' in \emph{33rd International Symposium on Software Testing and Analysis (ISSTA)}, 2024, pp. 819--831.

\bibitem{ThinkRepair}
X.~Yin, C.~Ni, S.~Wang, Z.~Li, L.~Zeng, and X.~Yang, ``Thinkrepair: Self-directed automated program repair,'' in \emph{33rd International Symposium on Software Testing and Analysis (ISSTA)}, 2024, pp. 1274--1286.

\bibitem{Contrastrepair}
J.~Kong, M.~Cheng, X.~Xie, S.~Liu, X.~Du, and Q.~Guo, ``Contrastrepair: Enhancing conversation-based automated program repair via contrastive test case pairs,'' \emph{arXiv preprint arXiv:2403.01971}, 2024.

\bibitem{SWEBench}
C.~E. Jimenez, J.~Yang, A.~Wettig, S.~Yao, K.~Pei, O.~Press, and K.~R. Narasimhan, ``{SWE}-bench: Can language models resolve real-world github issues?'' in \emph{12th International Conference on Learning Representations (ICLR)}, 2024.

\bibitem{AutoCodeRover}
Y.~Zhang, H.~Ruan, Z.~Fan, and A.~Roychoudhury, ``Autocoderover: Autonomous program improvement,'' in \emph{33rd International Symposium on Software Testing and Analysis (ISSTA)}, 2024, pp. 1592--1604.

\bibitem{SWE-agent}
J.~Yang, C.~Jimenez, A.~Wettig, K.~Lieret, S.~Yao, K.~Narasimhan, and O.~Press, ``Swe-agent: Agent-computer interfaces enable automated software engineering,'' \emph{Advances in Neural Information Processing Systems (NIPS)}, vol.~37, pp. 50\,528--50\,652, 2024.

\bibitem{Agentless}
C.~S. Xia, Y.~Deng, S.~Dunn, and L.~Zhang, ``Demystifying llm-based software engineering agents,'' \emph{Proc. ACM Softw. Eng. (FSE)}, 2025.

\bibitem{SpecRover}
H.~Ruan, Y.~Zhang, and A.~Roychoudhury, ``Specrover: Code intent extraction via llms,'' in \emph{47th International Conference on Software Engineering (ICSE)}, 2025, pp. 617--617.

\bibitem{LingmaSWE}
Y.~Ma, R.~Cao, Y.~Cao, Y.~Zhang, J.~Chen, Y.~Liu, Y.~Liu, B.~Li, F.~Huang, and Y.~Li, ``Lingma swe-gpt: An open development-process-centric language model for automated software improvement,'' in \emph{34th International Symposium on Software Testing and Analysis (ISSTA)}, 2025.

\bibitem{LingmaAgent}
Y.~Ma, Q.~Yang, R.~Cao, B.~Li, F.~Huang, and Y.~Li, ``Alibaba lingmaagent: Improving automated issue resolution via comprehensive repository exploration,'' \emph{arXiv preprint arXiv:2406.01422}, 2024.

\bibitem{SWEBenchM}
J.~Yang, C.~E. Jimenez, A.~L. Zhang, K.~Lieret, J.~Yang, X.~Wu, O.~Press, N.~Muennighoff, G.~Synnaeve, K.~R. Narasimhan \emph{et~al.}, ``Swe-bench multimodal: Do ai systems generalize to visual software domains?'' in \emph{13th International Conference on Learning Representations (ICLR)}, 2025.

\bibitem{Adversarial}
H.~Ye, A.~Z. Yang, C.~Hu, Y.~Wang, T.~Zhang, and C.~L. Goues, ``Adversarial reasoning for repair based on inferred program intent,'' \emph{arXiv preprint arXiv:2505.13008}, 2025.

\bibitem{Defects4J}
R.~Just, D.~Jalali, and M.~D. Ernst, ``Defects4j: A database of existing faults to enable controlled testing studies for java programs,'' in \emph{2014 International Symposium on Software Testing and Analysis (ISSTA)}, 2014, pp. 437--440.

\bibitem{LLM4APR_Jiang}
N.~Jiang, K.~Liu, T.~Lutellier, and L.~Tan, ``Impact of code language models on automated program repair,'' in \emph{45th International Conference on Software Engineering (ICSE)}, 2023, pp. 1430--1442.

\bibitem{BugReproduce}
R.~Cheng, M.~Tufano, J.~Cito, J.~Cambronero, P.~Rondon, R.~Wei, A.~Sun, and S.~Chandra, ``Agentic bug reproduction for effective automated program repair at google,'' \emph{arXiv preprint arXiv:2502.01821}, 2025.

\bibitem{GUIRepair_link}
\BIBentryALTinterwordspacing
GUIRepair, ``Guirepair,'' 2025. [Online]. Available: \url{https://sites.google.com/view/guirepair}
\BIBentrySTDinterwordspacing

\bibitem{AdbGPT}
S.~Feng and C.~Chen, ``Prompting is all you need: Automated android bug replay with large language models,'' in \emph{46th International Conference on Software Engineering (ICSE)}, 2024, pp. 1--13.

\bibitem{next-895}
\BIBentryALTinterwordspacing
AlibabaFusion/next-895, ``label should show in icononly mode,'' 2025. [Online]. Available: \url{https://github.com/alibaba-fusion/next/issues/894}
\BIBentrySTDinterwordspacing

\bibitem{AgentlessLite}
\BIBentryALTinterwordspacing
C.~S. Xia, Y.~Deng, S.~Dunn, and L.~Zhang, ``Agentless lite: Rag-based swe-bench software engineering scaffold,'' 2025. [Online]. Available: \url{https://github.com/sorendunn/Agentless-Lite}
\BIBentrySTDinterwordspacing

\bibitem{Pixelmatch}
\BIBentryALTinterwordspacing
pixelmatch, ``The smallest, simplest and fastest javascript pixel-level image comparison library,'' 2025. [Online]. Available: \url{https://github.com/mapbox/pixelmatch}
\BIBentrySTDinterwordspacing

\bibitem{Puppeteer}
\BIBentryALTinterwordspacing
Puppeteer, ``Javascript api for chrome and firefox,'' 2025. [Online]. Available: \url{https://github.com/puppeteer/puppeteer}
\BIBentrySTDinterwordspacing

\bibitem{chartjs}
\BIBentryALTinterwordspacing
Chart.js, ``Simple yet flexible javascript charting for designers and developers,'' 2025. [Online]. Available: \url{https://github.com/chartjs/Chart.js}
\BIBentrySTDinterwordspacing

\bibitem{openLayers}
\BIBentryALTinterwordspacing
OpenLayers, ``A high-performance, feature-packed library for all your mapping needs,'' 2025. [Online]. Available: \url{https://github.com/openlayers/openlayers}
\BIBentrySTDinterwordspacing

\bibitem{chartjs-10157}
\BIBentryALTinterwordspacing
chartjs/Chart.js 1509, ``borderradius gets ignored for the bottom corners of 0-value bars in bar chart when borderskipped and minbarlength are set,'' 2025. [Online]. Available: \url{https://github.com/chartjs/Chart.js/issues/10005}
\BIBentrySTDinterwordspacing

\bibitem{knowledgeInjection_Survey}
Z.~Song, B.~Yan, Y.~Liu, M.~Fang, M.~Li, R.~Yan, and X.~Chen, ``Injecting domain-specific knowledge into large language models: a comprehensive survey,'' \emph{arXiv preprint arXiv:2502.10708}, 2025.

\bibitem{Doc2Oracle}
S.~B. Hossain, R.~Taylor, and M.~Dwyer, ``Doc2oracle: Investigating the impact of javadoc comments on test oracle generation,'' \emph{arXiv preprint arXiv:2412.09360}, 2024.

\bibitem{GPT4o}
\BIBentryALTinterwordspacing
OpenAI, ``Gpt-4o,'' 2025. [Online]. Available: \url{https://platform.openai.com/docs/models/gpt-4o}
\BIBentrySTDinterwordspacing

\bibitem{RAPGen}
W.~Wang, Y.~Wang, S.~Joty, and S.~C. Hoi, ``Rap-gen: Retrieval-augmented patch generation with codet5 for automatic program repair,'' in \emph{31st ACM Joint European Software Engineering Conference and Symposium on the Foundations of Software Engineering (FSE)}, 2023, pp. 146--158.

\bibitem{text-embedding-3-small}
\BIBentryALTinterwordspacing
OpenAI, ``text-embedding-3-small,'' 2025. [Online]. Available: \url{https://platform.openai.com/docs/models/text-embedding-3-small}
\BIBentrySTDinterwordspacing

\bibitem{WebGen-Bench}
Z.~Lu, Y.~Yang, H.~Ren, H.~Hou, H.~Xiao, K.~Wang, W.~Shi, A.~Zhou, M.~Zhan, and H.~Li, ``Webgen-bench: Evaluating llms on generating interactive and functional websites from scratch,'' \emph{arXiv preprint arXiv:2505.03733}, 2025.

\bibitem{DCGen}
Y.~Wan, C.~Wang, Y.~Dong, W.~Wang, S.~Li, Y.~Huo, and M.~R. Lyu, ``Automatically generating ui code from screenshot: A divide-and-conquer-based approach,'' \emph{arXiv preprint arXiv:2406.16386}, 2024.

\bibitem{next-1509}
\BIBentryALTinterwordspacing
AlibabaFusion/next-1509, ``The popupcontainer configuration on configprovider does not take effect on dialog,'' 2025. [Online]. Available: \url{https://github.com/alibaba-fusion/next/issues/1508}
\BIBentrySTDinterwordspacing

\bibitem{SequenceR}
Z.~Chen, S.~Kommrusch, M.~Tufano, L.-N. Pouchet, D.~Poshyvanyk, and M.~Monperrus, ``Sequencer: Sequence-to-sequence learning for end-to-end program repair,'' \emph{IEEE Transactions on Software Engineering (TSE)}, vol.~47, no.~9, pp. 1943--1959, 2019.

\bibitem{RewardRepair}
H.~Ye, M.~Martinez, and M.~Monperrus, ``Neural program repair with execution-based backpropagation,'' in \emph{44th International Conference on Software Engineering (ICSE)}, 2022, pp. 1506--1518.

\bibitem{SelfAPR}
H.~Ye, M.~Martinez, X.~Luo, T.~Zhang, and M.~Monperrus, ``Selfapr: Self-supervised program repair with test execution diagnostics,'' in \emph{37th International Conference on Automated Software Engineering (ASE)}, 2022, pp. 1--13.

\bibitem{ITER}
H.~Ye and M.~Monperrus, ``Iter: Iterative neural repair for multi-location patches,'' in \emph{46th International Conference on Software Engineering (ICSE)}, 2024, pp. 1--13.

\bibitem{DLFix}
Y.~Li, S.~Wang, and T.~N. Nguyen, ``Dlfix: Context-based code transformation learning for automated program repair,'' in \emph{42nd International Conference on Software Engineering (ICSE)}, 2020, pp. 602--614.

\bibitem{DEAR}
------, ``Dear: A novel deep learning-based approach for automated program repair,'' in \emph{44th International Conference on Software Engineering (ICSE)}, 2022, pp. 511--523.

\bibitem{CoCoNut}
T.~Lutellier, H.~V. Pham, L.~Pang, Y.~Li, M.~Wei, and L.~Tan, ``Coconut: combining context-aware neural translation models using ensemble for program repair,'' in \emph{29th International Symposium on Software Testing and Analysis (ISSTA)}, 2020, pp. 101--114.

\bibitem{CURE}
N.~Jiang, T.~Lutellier, and L.~Tan, ``Cure: Code-aware neural machine translation for automatic program repair,'' in \emph{43rd International Conference on Software Engineering (ICSE)}, 2021, pp. 1161--1173.

\bibitem{KNOD}
N.~Jiang, T.~Lutellier, Y.~Lou, L.~Tan, D.~Goldwasser, and X.~Zhang, ``Knod: Domain knowledge distilled tree decoder for automated program repair,'' in \emph{45th International Conference on Software Engineering (ICSE)}, 2023, pp. 1251--1263.

\bibitem{Recoder}
Q.~Zhu, Z.~Sun, Y.-a. Xiao, W.~Zhang, K.~Yuan, Y.~Xiong, and L.~Zhang, ``A syntax-guided edit decoder for neural program repair,'' in \emph{29th ACM Joint European Software Engineering Conference and Symposium on the Foundations of Software Engineering (FSE)}, 2021, pp. 341--353.

\bibitem{Tare}
Q.~Zhu, Z.~Sun, W.~Zhang, Y.~Xiong, and L.~Zhang, ``Tare: Type-aware neural program repair,'' in \emph{45th International Conference on Software Engineering (ICSE)}, 2023, pp. 1443--1455.

\bibitem{NTR}
K.~Huang, J.~Zhang, X.~Meng, and Y.~Liu, ``Template-guided program repair in the era of large language models,'' in \emph{47th International Conference on Software Engineering (ICSE)}, 2025, pp. 367--379.

\bibitem{Repilot}
Y.~Wei, C.~S. Xia, and L.~Zhang, ``Copiloting the copilots: Fusing large language models with completion engines for automated program repair,'' in \emph{31st ACM Joint European Software Engineering Conference and Symposium on the Foundations of Software Engineering (FSE)}, 2023, pp. 172--184.

\bibitem{AlphaRepair}
C.~S. Xia and L.~Zhang, ``Less training, more repairing please: revisiting automated program repair via zero-shot learning,'' in \emph{30th ACM Joint European Software Engineering Conference and Symposium on the Foundations of Software Engineering (FSE)}, 2022, pp. 959--971.

\bibitem{FitRepair}
C.~S. Xia, Y.~Ding, and L.~Zhang, ``The plastic surgery hypothesis in the era of large language models,'' in \emph{38th International Conference on Automated Software Engineering (ASE)}, 2023, pp. 522--534.

\bibitem{SRepair}
J.~Xiang, X.~Xu, F.~Kong, M.~Wu, Z.~Zhang, H.~Zhang, and Y.~Zhang, ``How far can we go with practical function-level program repair?'' \emph{arXiv preprint arXiv:2404.12833}, 2024.

\bibitem{D4C}
J.~Xu, Y.~Fu, S.~H. Tan, and P.~He, ``Aligning the objective of llm-based program repair,'' \emph{arXiv preprint arXiv:2404.08877}, 2024.

\bibitem{LLM4APR_Fan}
Z.~Fan, X.~Gao, M.~Mirchev, A.~Roychoudhury, and S.~H. Tan, ``Automated repair of programs from large language models,'' in \emph{45th International Conference on Software Engineering (ICSE)}, 2023, pp. 1469--1481.

\bibitem{LLM4APR_Huang_ASE}
K.~Huang, X.~Meng, J.~Zhang, Y.~Liu, W.~Wang, S.~Li, and Y.~Zhang, ``An empirical study on fine-tuning large language models of code for automated program repair,'' in \emph{38th International Conference on Automated Software Engineering (ASE)}, 2023, pp. 1162--1174.

\bibitem{LLM4APR_Huang_TSE}
K.~Huang, J.~Zhang, X.~Bao, X.~Wang, and Y.~Liu, ``Comprehensive fine-tuning large language models of code for automated program repair,'' \emph{IEEE Transactions on Software Engineering (TSE)}, pp. 1--25, 2025.

\bibitem{LLM4APR_Wu}
Y.~Wu, N.~Jiang, H.~V. Pham, T.~Lutellier, J.~Davis, L.~Tan, P.~Babkin, and S.~Shah, ``How effective are neural networks for fixing security vulnerabilities,'' in \emph{32nd International Symposium on Software Testing and Analysis (ISSTA)}, 2023, pp. 1282--1294.

\bibitem{LLM4APR_Xia}
C.~S. Xia, Y.~Wei, and L.~Zhang, ``Automated program repair in the era of large pre-trained language models,'' in \emph{45th International Conference on Software Engineering (ICSE)}, 2023, pp. 1482--1494.

\bibitem{prism-1602}
\BIBentryALTinterwordspacing
Prism/prism-1602, ``Yaml strings fail with trailing comments,'' 2025. [Online]. Available: \url{https://github.com/PrismJS/prism/issues/1601}
\BIBentrySTDinterwordspacing

\bibitem{Leaderboard}
\BIBentryALTinterwordspacing
SWE-bench, ``Swe-bench multimodal leaderboard,'' 2025. [Online]. Available: \url{https://www.swebench.com/index.html#multimodal}
\BIBentrySTDinterwordspacing

\bibitem{ComputerUseAgent}
P.~Aggarwal and S.~Welleck, ``Programming with pixels: Computer-use meets software engineering,'' \emph{arXiv preprint arXiv:2502.18525}, 2025.

\bibitem{Playwright}
\BIBentryALTinterwordspacing
Playwright, ``Playwright enables reliable end-to-end testing for modern web apps,'' 2025. [Online]. Available: \url{https://playwright.dev/}
\BIBentrySTDinterwordspacing

\bibitem{Globant}
\BIBentryALTinterwordspacing
Globant, ``Globant code fixer agent,'' 2025. [Online]. Available: \url{https://ai.globant.com/}
\BIBentrySTDinterwordspacing

\bibitem{Zencoder}
\BIBentryALTinterwordspacing
Zencoder, ``The ai coding agent,'' 2025. [Online]. Available: \url{https://zencoder.ai/}
\BIBentrySTDinterwordspacing

\bibitem{GPT-4.1}
\BIBentryALTinterwordspacing
OpenAI, ``Gpt-4.1,'' 2025. [Online]. Available: \url{https://platform.openai.com/docs/models/gpt-4.1}
\BIBentrySTDinterwordspacing

\bibitem{o4-mini}
\BIBentryALTinterwordspacing
------, ``o4-mini,'' 2025. [Online]. Available: \url{https://platform.openai.com/docs/models/o4-mini}
\BIBentrySTDinterwordspacing

\bibitem{Resoning4APR}
Y.~Wei, O.~Duchenne, J.~Copet, Q.~Carbonneaux, L.~Zhang, D.~Fried, G.~Synnaeve, R.~Singh, and S.~I. Wang, ``Swe-rl: Advancing llm reasoning via reinforcement learning on open software evolution,'' \emph{arXiv preprint arXiv:2502.18449}, 2025.

\bibitem{ExecutionAgent}
I.~Bouzenia and M.~Pradel, ``You name it, i run it: An llm agent to execute tests of arbitrary projects,'' \emph{arXiv preprint arXiv:2412.10133}, 2024.

\bibitem{MoRepair}
B.~Yang, H.~Tian, J.~Ren, H.~Zhang, J.~Klein, T.~Bissyande, C.~Le~Goues, and S.~Jin, ``Morepair: Teaching llms to repair code via multi-objective fine-tuning,'' \emph{ACM Transactions on Software Engineering and Methodology (TOSEM)}, 2025.

\bibitem{RepairLlama}
A.~Silva, S.~Fang, and M.~Monperrus, ``Repairllama: Efficient representations and fine-tuned adapters for program repair,'' \emph{arXiv preprint arXiv:2312.15698}, 2023.

\bibitem{GAMMA}
Q.~Zhang, C.~Fang, T.~Zhang, B.~Yu, W.~Sun, and Z.~Chen, ``Gamma: Revisiting template-based automated program repair via mask prediction,'' in \emph{38th International Conference on Automated Software Engineering (ASE)}, 2023, pp. 535--547.

\bibitem{WILLIAMT}
H.~Zheng, I.~Shumailov, T.~Fan, A.~Hall, and M.~Payer, ``Fixing 7,400 bugs for 1\$: Cheap crash-site program repair,'' \emph{arXiv preprint arXiv:2505.13103}, 2025.

\bibitem{DRCodePilot}
J.~Zhao, D.~Yang, L.~Zhang, X.~Lian, Z.~Yang, and F.~Liu, ``Enhancing automated program repair with solution design,'' in \emph{39th International Conference on Automated Software Engineering (ASE)}, 2024, pp. 1706--1718.

\bibitem{CREF}
B.~Yang, H.~Tian, W.~Pian, H.~Yu, H.~Wang, J.~Klein, T.~F. Bissyand{\'e}, and S.~Jin, ``Cref: An llm-based conversational software repair framework for programming tutors,'' in \emph{33rd International Symposium on Software Testing and Analysis (ISSTA)}, 2024, pp. 882--894.

\bibitem{PredicateFix}
Y.~Xiao, W.~Wang, D.~Liu, J.~Zhou, S.~Cheng, and Y.~Xiong, ``Predicatefix: Repairing static analysis alerts with bridging predicates,'' \emph{arXiv preprint arXiv:2503.12205}, 2025.

\bibitem{TypeFix}
Y.~Peng, S.~Gao, C.~Gao, Y.~Huo, and M.~Lyu, ``Domain knowledge matters: Improving prompts with fix templates for repairing python type errors,'' in \emph{46th International Conference on Software Engineering (ICSE)}, 2024, pp. 1--13.

\bibitem{Openhands}
X.~Wang, B.~Li, Y.~Song, F.~F. Xu, X.~Tang, M.~Zhuge, J.~Pan, Y.~Song, B.~Li, J.~Singh \emph{et~al.}, ``Openhands: An open platform for ai software developers as generalist agents,'' in \emph{The Thirteenth International Conference on Learning Representations}, 2024.

\bibitem{Moatless}
A.~Antoniades, A.~{\"O}rwall, K.~Zhang, Y.~Xie, A.~Goyal, and W.~Wang, ``Swe-search: Enhancing software agents with monte carlo tree search and iterative refinement,'' \emph{arXiv preprint arXiv:2410.20285}, 2024.

\bibitem{Marscode}
Y.~Liu, P.~Gao, X.~Wang, J.~Liu, Y.~Shi, Z.~Zhang, and C.~Peng, ``Marscode agent: Ai-native automated bug fixing,'' \emph{arXiv preprint arXiv:2409.00899}, 2024.

\bibitem{APRAgent_Study}
X.~Meng, Z.~Ma, P.~Gao, and C.~Peng, ``An empirical study on llm-based agents for automated bug fixing,'' \emph{arXiv preprint arXiv:2411.10213}, 2024.

\bibitem{Designrepair}
M.~Yuan, J.~Chen, Z.~Xing, A.~Quigley, Y.~Luo, T.~Luo, G.~Mohammadi, Q.~Lu, and L.~Zhu, ``Designrepair: Dual-stream design guideline-aware frontend repair with large language models,'' \emph{arXiv preprint arXiv:2411.01606}, 2024.

\bibitem{Iris}
Y.~Zhang, S.~Chen, L.~Fan, C.~Chen, and X.~Li, ``Automated and context-aware repair of color-related accessibility issues for android apps,'' in \emph{31st ACM Joint European Software Engineering Conference and Symposium on the Foundations of Software Engineering (FSE)}, 2023, pp. 1255--1267.

\bibitem{LLM4GUI_Survey}
F.~Tang, H.~Xu, H.~Zhang, S.~Chen, X.~Wu, Y.~Shen, W.~Zhang, G.~Hou, Z.~Tan, Y.~Yan \emph{et~al.}, ``A survey on (m) llm-based gui agents,'' \emph{arXiv preprint arXiv:2504.13865}, 2025.

\bibitem{LLM4GUIAPP_Survey}
S.~Yu, C.~Fang, Z.~Tuo, Q.~Zhang, C.~Chen, Z.~Chen, and Z.~Su, ``Vision-based mobile app gui testing: A survey,'' \emph{arXiv preprint arXiv:2310.13518}, 2023.

\bibitem{GPTDroid}
Z.~Liu, C.~Chen, J.~Wang, M.~Chen, B.~Wu, X.~Che, D.~Wang, and Q.~Wang, ``Make llm a testing expert: Bringing human-like interaction to mobile gui testing via functionality-aware decisions,'' in \emph{46th International Conference on Software Engineering (ICSE)}, 2024, pp. 1--13.

\bibitem{SoapTesting}
Y.~Su, Z.~Xing, C.~Wang, C.~Chen, X.~Xu, Q.~Lu, and L.~Zhu, ``Automated soap opera testing directed by llms and scenario knowledge: Feasibility, challenges, and road ahead,'' \emph{arXiv preprint arXiv:2412.08581}, 2024.

\bibitem{Report2Test_Fazz}
M.~Fazzini, M.~Prammer, M.~d'Amorim, and A.~Orso, ``Automatically translating bug reports into test cases for mobile apps,'' in \emph{27th International Symposium on Software Testing and Analysis (ISSTA)}, 2018, pp. 141--152.

\bibitem{Report2Test_Zhao}
Y.~Zhao, T.~Yu, T.~Su, Y.~Liu, W.~Zheng, J.~Zhang, and W.~G. Halfond, ``Recdroid: automatically reproducing android application crashes from bug reports,'' in \emph{41st International Conference on Software Engineering (ICSE)}, 2019, pp. 128--139.

\bibitem{Video2Test_Ber}
C.~Bernal-C{\'a}rdenas, N.~Cooper, K.~Moran, O.~Chaparro, A.~Marcus, and D.~Poshyvanyk, ``Translating video recordings of mobile app usages into replayable scenarios,'' in \emph{42nd International Conference on Software Engineering (ICSE)}, 2020, pp. 309--321.

\bibitem{Video2Test_Hav}
M.~Havranek, C.~Bernal-C{\'a}rdenas, N.~Cooper, O.~Chaparro, D.~Poshyvanyk, and K.~Moran, ``V2s: A tool for translating video recordings of mobile app usages into replayable scenarios,'' in \emph{43rd International Conference on Software Engineering: Companion Proceedings (ICSE-Companion)}, 2021, pp. 65--68.

\bibitem{SoapTesting_Report}
Y.~Su, Z.~Han, Z.~Xing, X.~Xia, X.~Xu, L.~Zhu, and Q.~Lu, ``Constructing a system knowledge graph of user tasks and failures from bug reports to support soap opera testing,'' in \emph{37th International Conference on Automated Software Engineering (ASE)}, 2022, pp. 1--13.

\bibitem{SoapOperaTg}
Y.~Su, Z.~Han, Z.~Xing, X.~Xu, L.~Zhu, and Q.~Lu, ``Soapoperatg: A tool for system knowledge graph based soap opera test generation,'' in \emph{45th International Conference on Software Engineering: Companion Proceedings (ICSE-Companion)}, 2023, pp. 51--54.

\end{thebibliography}

\end{document}